\documentclass[useAMS,usenatbib]{mn2e}
\usepackage{epsf}
\usepackage{epsfig}
\usepackage{subfigure}

\newcommand{\refsim}{\textsc{ref}}
\newcommand{\agn}{\textsc{agn}}
\newcommand{\nocool}{\textsc{nocool}}

\newcommand{\planck}{\textit{Planck}}
\newcommand{\wmap}{\textit{WMAP}7}

\title[{tSZ power spectrum and \planck}]{The thermal Sunyaev Zel'dovich effect power spectrum in light of \planck}

\author[I.~G.~McCarthy et~al.]{I.~G.~McCarthy$^1$\thanks{E-mail:
i.g.mccarthy@ljmu.ac.uk}, A.~M.~C.~Le Brun$^1$, J. Schaye$^2$, G. P. Holder$^3$\\
$^{1}$Astrophysics Research Institute, Liverpool John Moores University, 146 Brownlow Hill, Liverpool L3 5RF\\
$^{2}$Leiden Observatory, Leiden University, P. O. Box 9513, 2300 RA Leiden, the Netherlands\\
$^{3}$Department of Physics, McGill University, 3600 Rue University, Montreal, Quebec H3A 2T8, Canada
}

\begin{document}

\date{Accepted ... Received ...}

\pagerange{\pageref{firstpage}--\pageref{lastpage}} \pubyear{2014}

\maketitle

\label{firstpage}

\begin{abstract}
The amplitude of the thermal Sunyaev Zel'dovich effect (tSZ) power spectrum is extremely sensitive to the abundance of the most massive dark matter haloes (galaxy clusters) and therefore to fundamental cosmological parameters that control their growth, such as $\sigma_8$ and $\Omega_m$.  Here we explore the sensitivity of the tSZ power spectrum to important non-gravitational (`sub-grid') physics by employing the cosmo-OWLS suite of large-volume cosmological hydrodynamical simulations, run in both the \planck~and \wmap~best-fit cosmologies.  On intermediate and small angular scales ($\ell \ga 1000$, or $\theta \la 10$ arcmin), accessible with the South Pole Telescope (SPT) and the Atacama Cosmology Telescope (ACT), the predicted tSZ power spectrum is highly model dependent, with gas ejection due to Active Galactic Nuclei (AGN) feedback having a particularly large effect.  However, at large scales, observable with the \planck~telescope, the effects of sub-grid physics are minor.  Comparing the simulated tSZ power spectra with observations, we find a significant amplitude offset on all measured angular scales (including large scales), if the \planck~best-fit cosmology is assumed by the simulations.  This is shown to be a generic result for all current models of the tSZ power spectrum.  By contrast, if the \wmap~cosmology is adopted, there is full consistency with the \planck~tSZ power spectrum measurements on large scales and agreement at the 2 sigma level with the SPT and ACT measurements at intermediate scales for our fiducial AGN model, which \citet{LeBrun2014} have shown reproduces the `resolved' properties of the local group and cluster population remarkably well.  These findings strongly suggest that there are significantly fewer massive galaxy clusters than expected for the \planck~best-fit cosmology, which is consistent with recent measurements of the tSZ number counts.  Our findings therefore pose a significant challenge to the cosmological parameter values preferred (and/or the model adopted) by the Planck primary CMB analyses.
\end{abstract}


\section{Introduction}

The hot gas in galaxy groups and clusters, called the intracluster medium (ICM), acts as a secondary source of anisotropies in the cosmic microwave background (CMB).  CMB photons passing through a cluster are on average likely to inverse Compton scatter off hot electrons in the ICM, which gives the photons a small energy kick.  This produces a slight intensity/temperature decrement at radio wavelengths and a slight increment at millimeter wavelengths, known as the thermal Sunyaev-Zel'dovich effect (hereafter, tSZ; \citealt{Sunyaev1972}, see \citealt{Birkinshaw1999} for a review).  If the cluster is moving with respect to the CMB rest frame, an additional distortion of the CMB due to the Doppler effect will also be produced, known as the kinetic Sunyaev-Zel'dovich effect (kSZ).  The kSZ is significantly weaker than the tSZ, except near the tSZ null at $\approx 218$ GHz (i.e., the frequency at which the number of photons scattered up from lower energies cancels the number of photons being scattered up to higher energies).  For the present study, we concern ourselves with the tSZ only, noting that the kSZ signal has been detected for the first time only very recently (e.g., \citealt{Hand2012,Sayers2013}).

The tSZ signal on the sky is highly sensitive to the fundamental cosmological parameters that control the growth of galaxy clusters (e.g., \citealt{Carlstrom2002,Komatsu2002}), offering an important and independent measurement of parameters such as $\sigma_8$ and $\Omega_m$ (which can be constrained through the tSZ power spectrum amplitude and tSZ cluster number counts), as well as $H_0$ (by exploiting the differing dependencies of the tSZ and X-ray signals of the ICM to measure a physical size of clusters independent of their redshift) and a tool to test models of the evolution of dark energy (e.g., by measuring the redshift evolution of the number counts).  It is therefore unsurprising that there are large numbers of tSZ surveys in the works (e.g., ACT, SPT, \planck, APEX-SZ, MUSTANG, CARMA, ALMA, AMI, AMiBA).

Its use as a cosmological probe is, however, complicated by the fact that the tSZ signal is sensitive to the astrophysics governing the thermal state of the ICM, since the magnitude of the tSZ depends directly on the (line of sight integral of) pressure of the hot gas.  The pressure, in turn, is set by the depth of the dark matter potential well and the entropy of the hot gas, which can be significantly altered by non-gravitational processes such as radiative cooling and feedback from processes related to galaxy formation (e.g., \citealt{Voit2005,Nagai2007,McCarthy2011}).  Indeed, recent studies have shown that the tSZ power spectrum is sensitive to ICM modelling details on scales of a few arcminutes (e.g., \citealt{Holder2007,Shaw2010,Battaglia2010,Trac2011}) where, until recently, tSZ power spectrum constraints have been limited to.

However, it is noteworthy that important progress has been made in recent years on modelling the effects of cooling and feedback on the ICM, so much so that reasonably realistic populations of clusters, which match a wide variety of observed properties, are now being produced in cosmological simulations (e.g., \citealt{Bower2008,Puchwein2008,Short2009,McCarthy2010,McCarthy2011,Planelles2013,LeBrun2014}).  Furthermore, measurements of the tSZ power spectrum are now being made on larger angular scales (of a few degrees) with the \planck~telescope \citep{Planck2013c}, which are significantly less sensitive to uncertain baryonic physics (e.g., \citealt{Komatsu1999}).  This should give a renewed emphasis on the tSZ as a cosmological probe.

In the present study, we take state-of-the-art cosmological hydrodynamical simulations and construct large simulated tSZ skies and make comparisons with the latest `unresolved' (power spectrum) tSZ measurements from the \planck~telescope, as well as from the South Pole Telescope (SPT), and the Atacama Cosmology Telescope (ACT).  From this comparison we arrive at the robust conclusion that there is a significant tension between existing tSZ power spectrum measurements and the cosmological parameter values preferred (and/or the model adopted) by the Planck primary CMB analyses \citep{Planck2013a}.

The present study is organised as follows.  In Section 2 we briefly describe the cosmo-OWLS simulation suite used here and our mapmaking procedure.  \citet{LeBrun2014} have compared these simulations with the observed properties of local groups and clusters and concluded that the fiducial Active Galactic Nuclei (AGN) feedback model performs remarkably well, reproducing the observed trends over a wide range of halo masses and radii.  In Section 3 we compare the predicted pressure distributions of the simulated groups and clusters with observations of local systems.  In Section 4 we dissect the theoretical tSZ power spectra into its contributions from hot gas in haloes in bins of mass, redshift, and radius.  In Section 5 we compare the predicted tSZ power spectra with observations.  In Section 6 we compare our predicted tSZ power spectra with those of other models.  Finally, in Section 7 we summarize and discuss our findings.

\section{Simulations}

We employ the cosmo-OWLS suite of cosmological hydrodynamical simulations described in detail in \citet{LeBrun2014} (hereafter, L14; see also \citealt{vanDaalen2013}).  cosmo-OWLS is an extension of the OverWhelmingly Large Simulations project (OWLS; \citealt{Schaye2010}) designed with cluster cosmology and large scale-structure surveys in mind.  The cosmo-OWLS suite consists of large-volume, $(400 \ h^{-1} \ {\rm Mpc})^3$, periodic box hydrodynamical simulations with $1024^3$ baryon and dark matter particles (each) and with updated initial conditions based either on the maximum-likelihood cosmological parameters derived from the 7-year {\it WMAP} data (hereafter \wmap; \citealt{Komatsu2011}) \{$\Omega_{m}$, $\Omega_{b}$, $\Omega_{\Lambda}$, $\sigma_{8}$, $n_{s}$, $h$\} = \{0.272, 0.0455, 0.728, 0.81, 0.967, 0.704\} or the \planck~data \citep{Planck2013a} = \{0.3175, 0.0490, 0.6825, 0.834, 0.9624, 0.6711\}.  This yields dark matter and (initial) baryon particle masses of $\approx4.44\times10^{9}~h^{-1}~\textrm{M}_{\odot}$ ($\approx3.75\times10^{9}~h^{-1}~\textrm{M}_{\odot}$) and $\approx8.12\times10^{8}~h^{-1}~\textrm{M}_{\odot}$ ($\approx7.54\times10^{8}~h^{-1}~\textrm{M}_{\odot}$) for the \planck~(\wmap) cosmology.  The extension to large volumes is quite important for the present study, since the tSZ power spectrum is dominated by massive dark matter haloes with $M \sim 10^{14} ~\textrm{M}_{\odot}$, which have very low space densities of $\sim 10^{-5}$ Mpc$^{-3}$ (e.g., \citealt{Jenkins2001}).

As in OWLS, the comoving gravitational softening lengths for the baryon and dark matter particles are set to $1/25$ of the initial mean inter-particle spacing (e.g., \citealt{Mo2010}) but are limited to a maximum physical scale of $4~h^{-1}$ kpc (Plummer equivalent). The switch from a fixed comoving to a fixed proper softening happens at $z = 2.91$.  (Note that current measurements of the tSZ power spectrum probe physical scales that are two to three orders of magnitude larger than the gravitational softening of our simulations.)  We use $N_{\rm ngb} = 48$ neighbours for the SPH interpolation and the minimum SPH smoothing length is limited to $0.01$ of the gravitational softening.  

\begin{table*}
\centering
\caption{cosmo-OWLS runs presented here and their included sub-grid physics.  Each model has been run in both the \wmap~and \planck~cosmologies.}
\begin{tabular}{|l|l|l|l|l|l|l|}
         \hline
	Simulation & UV/X-ray background & Cooling & Star formation & SN feedback & AGN feedback & $\Delta T_{\rm heat}$ \\
	\hline
        \nocool & Yes & No & No & No & No & ...\\
        \refsim & Yes & Yes & Yes & Yes & No & ...\\
        \agn~8.0 & Yes & Yes & Yes & Yes & Yes & $10^{8.0}$ K\\
        \agn~8.5 & Yes & Yes & Yes & Yes & Yes & $10^{8.5}$ K\\
        \agn~8.7 & Yes & Yes & Yes & Yes & Yes & $10^{8.7}$ K\\
        \hline
\end{tabular}
\label{table:cosmo_owls}
\end{table*}

The simulations were run using a version of the Lagrangian TreePM-SPH code \textsc{gadget3} \citep[last described in][]{Springel2005b}, which was significantly modified to include new `sub-grid' physics as part of the OWLS project.  Starting from identical initial conditions (for a given cosmology), key parameters controlling the nature and strength of feedback are systematically varied.  As in L14, we use five different physical models: \nocool, \refsim, \agn~8.0, \agn~8.5, and \agn~8.7.  The \nocool~model is a standard non-radiative (`adiabatic') model.  \refsim~is the OWLS reference model, which includes sub-grid prescriptions for star formation \citep{Schaye2008}, metal-dependent radiative cooling \citep{Wiersma2009a}, stellar evolution, mass loss, and chemical enrichment \citep{Wiersma2009b}, and a kinetic supernova feedback prescription \citep{DallaVecchia2008}.

The three AGN models (\agn~8.0, \agn~8.5, and \agn~8.7) include the same sub-grid prescriptions as the \refsim~model, but also include a prescription for black hole growth and feedback from active galactic nuclei (\citealt{Booth2009}, a modified version of the model developed originally by \citealt{Springel2005a}).  The black holes store up enough energy\footnote{As in \citet{Booth2009} we use 1.5\% of the rest mass energy of accreted gas for feedback.  This efficiency choice results in a reasonable match to the normalisation of the local black hole scaling relations (\citealt{Booth2009}, see also L14) and is insensitive to the precise value of $\Delta T_{\rm heat}$.} until they are able to raise the temperature of neighboring gas by a pre-defined level, $\Delta T_{\rm heat}$.  The three AGN models differ only in their choice of the heating temperature $\Delta T_{\rm heat}$, which is the most critical parameter of the AGN feedback model\footnote{We note that the \agn~8.0 model was referred to as `AGN' in previous OWLS papers and was studied in \citet{McCarthy2010,McCarthy2011} with a $100 \ h^{-1}$ Mpc box, a {\it WMAP}3 cosmology and $8\times$ smaller particle mass.}.  Note that since the same amount of gas is being heated in these models, more time is required for the BHs to accrete enough mass to be able to heat neighbouring gas to a higher temperature.  Thus, increasing the heating temperature leads to more {\it bursty} and more violent feedback.

Table 1 provides a list of the runs used here and the sub-grid physics that they include.  In Appendix A we present a resolution study, concluding that our simulations are reasonably well converged.

\subsection{Thermal SZ effect (tSZ) maps}
 
The magnitude of the tSZ is set by the dimensionless Compton $y$ parameter, defined as:

\begin{equation}
y \equiv \int \sigma_T \frac{k_b T}{m_e c^2} n_e dl \ \ \ ,
\end{equation}

\noindent where $\sigma_T$ is the Thomson cross-section, $k_B$ is Boltzmann's constant, $T$ is the gas temperature, $m_e$ is the electron rest mass, $c$ is the speed of light, and $n_e$ is the electron number density.  Thus, $y$ is proportional to the electron pressure integrated along the observer's line of the sight, back to the epoch of reionization.

To produce Compton $y$ maps, we stack randomly rotated and translated snapshots at differing redshifts along the line of sight \citep{daSilva2000} back to $z=3$.  (This is sufficiently high redshift for approximate convergence in the tSZ power spectrum; see Fig.~\ref{fig:cuts}.)  We follow the approach of \citet{Roncarelli2006,Roncarelli2007} and calculate the quantity

\begin{equation}
\Upsilon_i \equiv \sigma_T \frac{k_b T_i}{m_e c^2} \frac{m_i}{\mu_{e,i} m_H}
\end{equation}

\noindent for the $i^{\rm th}$ gas particle. Here $T_i$ is the temperature of the gas particle, $m_i$ is the gas particle mass, $\mu_{e,i}$ is the mean molecular weight per free electron of the gas particle (which depends on its metallicity), and $m_H$ is the atomic mass of hydrogen.  Note that $\Upsilon_i$ has dimensions of area.

The total contribution to the Compton $y$ parameter in a given pixel by the $i^{\rm th}$ particle is obtained by dividing $\Upsilon_i$ by the physical area of the pixel at the angular diameter distance of the particle from the observer; i.e., $y_i \equiv \Upsilon_i/L_{pix,i}^2$.  We adopt an angular pixel size of 2.5 arcsec, which is better than what can be achieved with current tSZ instrumentation but is similar to the spatial resolution of X-ray telescopes like {\it Chandra}.  We opt for this high angular resolution because we are producing X-ray maps simultaneously with the tSZ maps.

Finally, we smooth $y_i$ onto the map using the SPH smoothing kernel, adopting as the smoothing length the 3D physical smoothing length of the particle (calculated by \textsc{gadget3}) divided by the angular diameter distance of the particle; i.e., the {\it angular} extent of the particle's smoothing length.   We have verified that the exact choice of smoothing kernel or smoothing length is inconsequential for the tSZ power spectrum over the range of angular scales considered here ($\ell < 10000$, corresponding to $\theta \ga 1$ arcmin), by comparing the power spectrum produced using the fiducial SPH-smoothed maps with that produced from maps generated using a simple `nearest-grid point' method (they are virtually identical).

\begin{figure*}
\includegraphics[width=\textwidth]{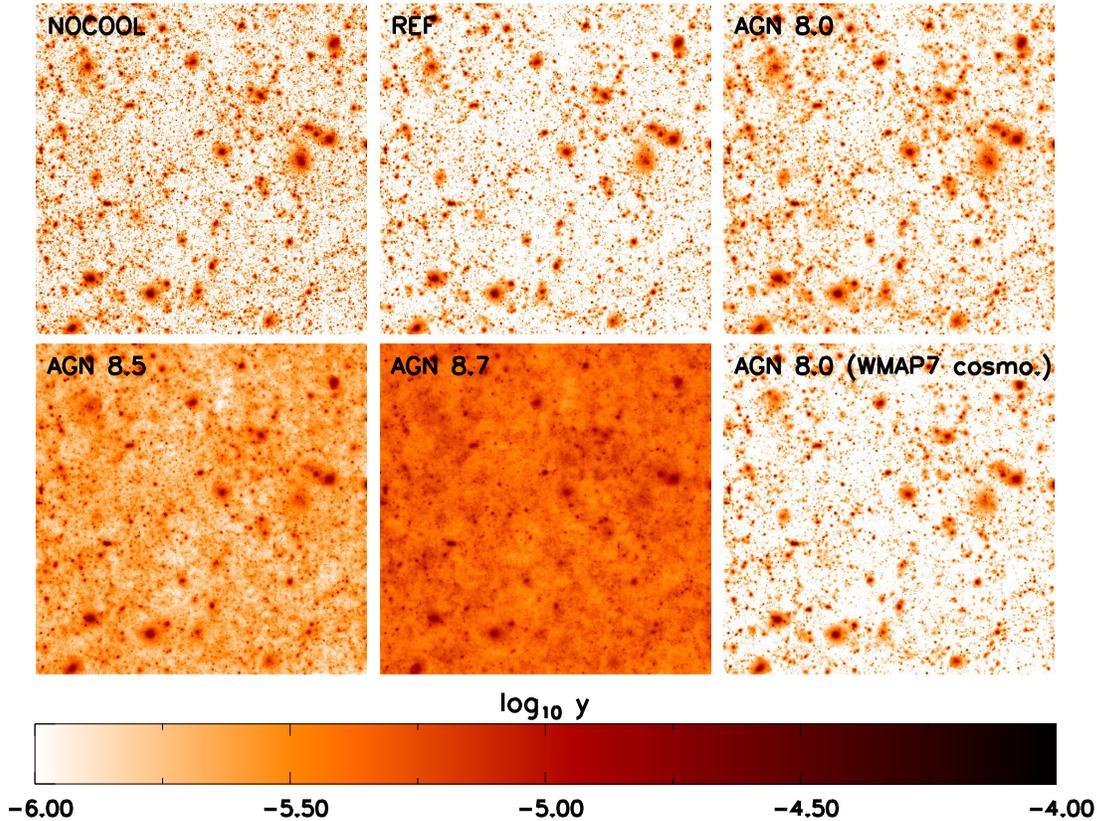}
\caption{\label{fig:maps}
Example simulated Compton-y maps for the five different physical models in the \planck~cosmology, along with one in the \wmap~cosmology (bottom right panel).  Each map is $5\deg \times 5\deg$ and adopts the same viewing angle (i.e., the same randomly-selected rotations and translations are applied in each case).  Differences between the five \planck~cosmology maps are due entirely to differences in the sub-grid physics, with gas ejection associated with AGN feedback having a particularly large effect.  For a fixed physical model, the difference between the \planck~and the \wmap~cosmology is also readily visible, with more (and larger) systems present in the \planck~cosmology run.
}
\end{figure*}

Previous studies found that cosmic variance can be an issue for the tSZ power spectrum calculated from maps produced from self-consistent cosmological hydrodynamical simulations, due to their finite box size and therefore limited field of view (e.g., \citealt{White2002,Battaglia2010}).  Indeed, most previous simulation studies produced maps of only a few square degrees, while current observational surveys being conducted are hundreds of square degrees.  Our larger simulations allow us to produce larger maps of $5\deg \times 5\deg$ ($7200 \times 7200$ pixels), but cosmic variance is still an issue.  We therefore produce 10 maps corresponding to different viewing angles (by randomly rotating and translating the boxes) for each simulation.  We note that $5\deg$ corresponds approximately to the co-moving length of the simulation box ($400 \ h^{-1}$ Mpc) at $z=3$.  Thus, at high redshift the 10 maps will probe many of the same structures.  At lower redshifts (which dominate the tSZ power spectrum, as we show in Section 4), however, the 25 $\deg^2$ field of view occupies only a relatively small fraction of the simulated volume, and therefore the maps are effectively independent.  In Appendix B we show the map-to-map scatter around the mean and median tSZ power spectra.

As an example, we show in Fig.~\ref{fig:maps} simulated Compton $y$ maps for the five different physical models in the \planck~cosmology, along with one in the \wmap~cosmology.  All maps adopt the same viewing angle (i.e., the same randomly-selected rotations and translations are applied in each case).  Thus, the differences between the 5 \planck~cosmology maps are due entirely to differences in the sub-grid physics.  Particularly noticeable is the impact of AGN feedback, which ejects gas from dark matter haloes out into the intergalactic medium.  The two right most panels compare the same physical model (the fiducial AGN model, \agn~8.0) in the two different cosmologies.  The cosmological and astrophysical dependencies of the tSZ signal are easily visible by eye in Fig.~\ref{fig:maps}.

For each simulation we compute the tSZ angular power spectrum by averaging over the power spectra computed for each of the 10 maps.

In addition to tSZ maps, we also create halo catalogues for our light cones using a standard friends-of-friends algorithm run on the snapshot data.  In Section 4 we use the halo catalogues to deconstruct the theoretical power spectra into its contributions from haloes of different mass and redshifts and from different radial ranges within the haloes.

\section{Pressure profiles}

\begin{figure*}
\includegraphics[width=0.49\textwidth]{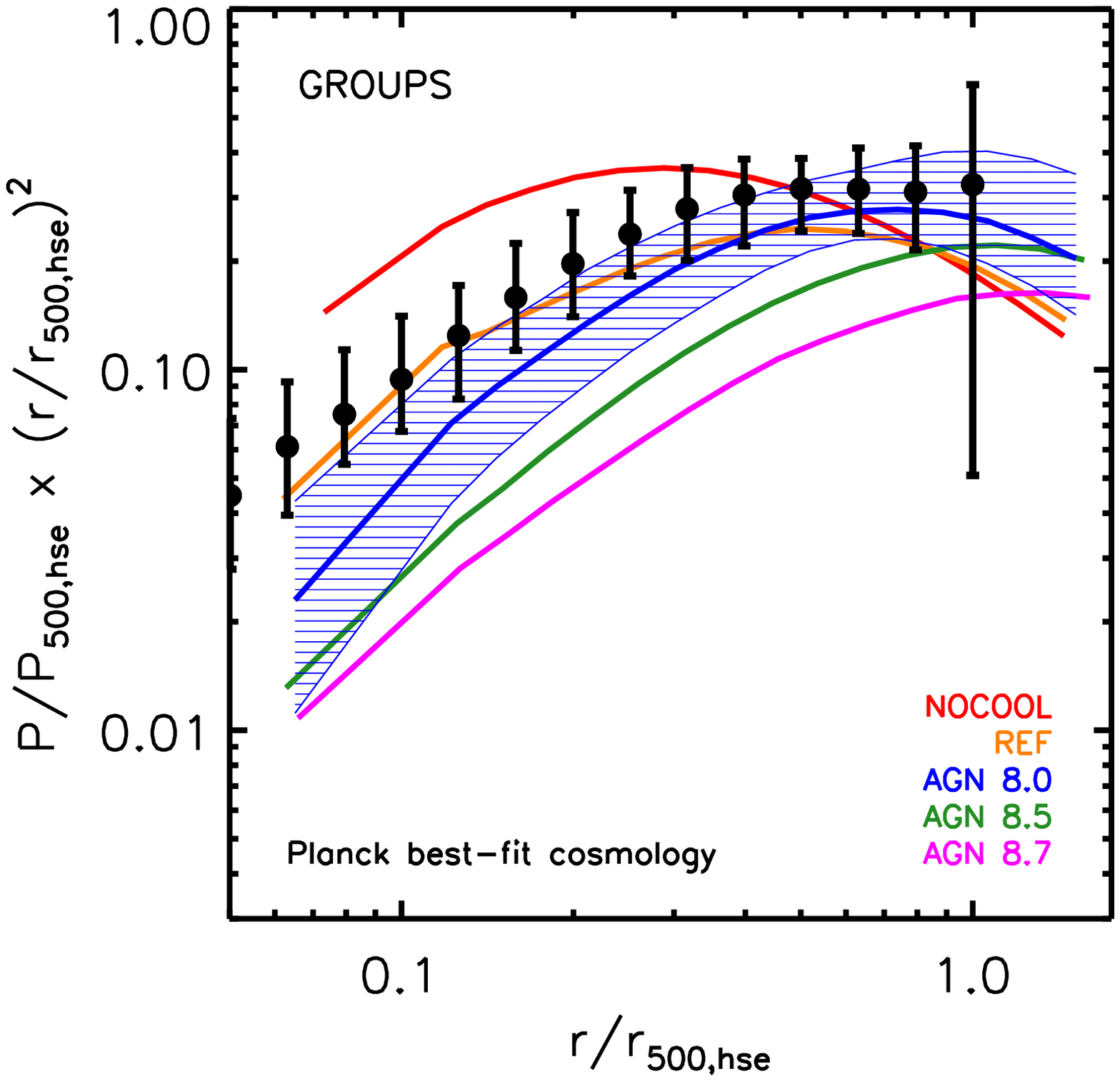}
\includegraphics[width=0.49\textwidth]{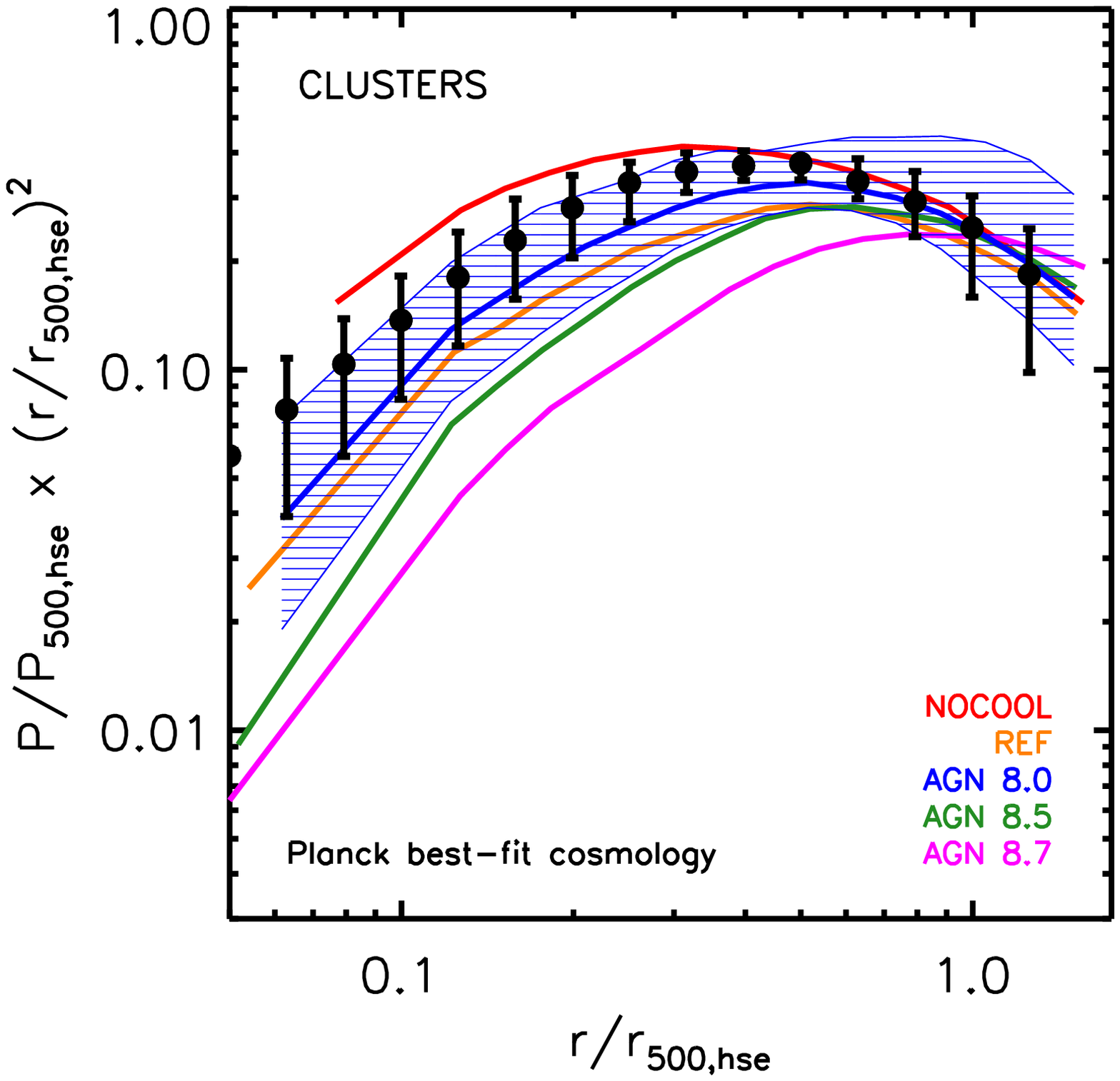}
\caption{\label{fig:pressure_profs}
Radial electron pressure profiles of groups ($13 \la \log[M_{500,hse}h/~\textrm{M}_{\odot}] \la 14.1$; {\emph left}) and clusters ($13.8 \la \log[M_{500,hse}h/~\textrm{M}_{\odot}] \la 14.8$; {\emph right}) at $z \approx 0$. The filled black circles with error bars correspond to the observational data of \citet{Sun2011} ({\emph left}, groups) and the REXCESS sample of \citet{Bohringer2007} and \citet{Arnaud2010} ({\emph right}, clusters).  The error bars enclose 68 per cent of the observed systems.  The curves represent medians of the different simulations, with the shaded region enclosing 68 per cent of the simulated systems for \agn~8.0 model.  The fiducial AGN model (\agn~8.0), which reproduces the local X-ray and optical scaling relations best (L14), reproduces the observed pressure profiles of groups (outside $\ga 0.3 r_{500}$) and clusters well.  }
\end{figure*}

Before proceeding to an analysis of the tSZ power spectrum, we first briefly (re-)examine the degree of realism of the five physical models by comparing to the observed properties of local X-ray-selected galaxy clusters.  We note that L14 have already subjected these models to a full battery of observational tests at low redshift, including global X-ray, tSZ, optical, and BH scaling relations.  One of the conclusions of that study is that the fiducial AGN model (\agn~8.0) reproduces virtually all of the observed local relations reasonably well (including their intrinsic scatter), while models that neglect AGN feedback (\refsim) suffer from significant overcooling, producing a factor of 3-5 times more mass in stars than observed.  AGN models with increased heating temperatures (particularly \agn~8.7), on the other hand, eject too much gas from the progenitors of groups and clusters, yielding present-day groups and clusters with lower gas mass fractions and higher entropy than observed (it is the low-entropy gas that is preferentially heated and ejected, at high redshift).  An important caveat to bear in mind, however, is that the role of observational selection is not yet well understood and this currently limits our ability to perform detailed quantitative comparisons between the models and observations (see discussion in L14).

Of direct relevance for the tSZ angular power spectrum is the electron pressure distribution of the hot gas and its dependence on halo mass and redshift, which was not examined in L14.  To make a like-with-like comparison to the data, we construct synthetic X-ray observations and derive the gas density and temperature (and therefore pressure) by fitting to synthetic spatially-resolved X-ray spectra (see L14 for details).  We use the same synthetic observations and assume hydrostatic equilibrium to `measure' the mass, $M_{500,hse}$, and the corresponding overdensity radius $r_{500,hse}$ for each of the simulated clusters.

In Fig.~\ref{fig:pressure_profs} we plot the radial electron pressure profiles of $z=0$ groups (left panel) and clusters (right panel) for the various simulations and compare to X-ray observations of local, bright X-ray systems.  For the observations, we compare to the {\it Chandra} group sample of \citet{Sun2011}, while for the clusters we compare to the {\it XMM-Newton} REXCESS sample \citep{Bohringer2007}.  Note that this is the same data from which \citet{Arnaud2010} derived the `universal pressure profile', which adopts a generalised NFW form (see \citealt{Nagai2007}).  Instead of plotting the universal pressure profile, we plot the best-fit profiles for the individual REXCESS systems (i.e., with system-to-system scatter included).  

We normalise the radial coordinate by $r_{500,hse}$, the radius within which the mean mass density is 500 times the critical density for closure (which is typically the radius out to which good quality X-ray data can presently probe).  We normalise the electron pressure by the `virial pressure' $P_{500,hse} \equiv n_{e,500} k_B T_{500,hse}$, where $k_B T_{500,hse} \equiv \mu m_p G M_{500,hse} / 2 r_{500,hse}$ is the virial temperature and $n_{e,500}$ is the mean electron density within $r_{500,hse}$ assuming the universal baryon fraction $f_b \equiv \Omega_b/\Omega_m$; i.e., $n_{e,500} \equiv 500 f_b \rho_{crit}(z) / \mu_e m_H$.  To reduce the dynamic range on the y-axis further, we scale the normalised pressure by a factor $(r/r_{500,hse})^2$ for both simulations and observations.  We also scale the observed pressure profiles to our adopted cosmology when comparing to the simulations (noting particularly that $P_{500,hse}$ depends on the adopted $f_b$).  Lastly, as the shape and amplitude of the pressure profiles are fairly strong functions of halo mass, we have re-sampled the simulated cluster mass distribution in order to achieve approximately the same median mass as the observed samples.  In particular, for the groups we select systems in the (true) mass range $5.8\times10^{13}~\textrm{M}_{\odot} < M_{500} < 1.5\times10^{14}~\textrm{M}_{\odot}$ to achieve a median mass of $M_{500,hse} \approx 8.6\times10^{13}~\textrm{M}_{\odot}$,  while for the cluster comparison we select systems in the mass range $2.5\times10^{14}~\textrm{M}_{\odot} < M_{500} < 10^{15}~\textrm{M}_{\odot}$ to achieve a median mass of $M_{500,hse} \approx 3.5\times10^{14}~\textrm{M}_{\odot}$.

From Fig.~\ref{fig:pressure_profs} it is immediately apparent that the pressure distribution of the hot gas is strongly model dependent, with large differences between the models within $\sim r_{500,hse}$ for groups and $\sim 0.5 r_{500,hse}$ for clusters.  At $\sim 0.1 r_{500,hse}$, for example, the pressure can vary by up to an order of magnitude from model to model.  The tSZ power spectrum at currently accessible angular scales is sensitive to intermediate radii (see Fig.~\ref{fig:cuts} below), implying we should expect some sensitivity to non-gravitational physics.

Consistent with L14, we find that the fiducial AGN model (\agn~8.0) appears to perform best.  The \refsim~model, which neglects AGN feedback, performs similarly well, but at the expense of significant overcooling; i.e., too high stellar masses (not shown here, see L14).  Increasing the AGN heating temperature leads to a strong suppression of the gas density at small and intermediate radii, which in turn yields electron pressures that are significantly lower than observed.  However, it is important to bear in mind that at present we can only make these kind of comparisons for {\it local} groups and clusters, where the data quality is sufficiently high.  The tSZ power spectrum, however, has a non-negligible contribution from high redshift clusters (out to $z \sim 1.5$) and it is unclear which (if any) of the models performs reasonably well there.

\begin{figure}
\includegraphics[width=0.995\columnwidth]{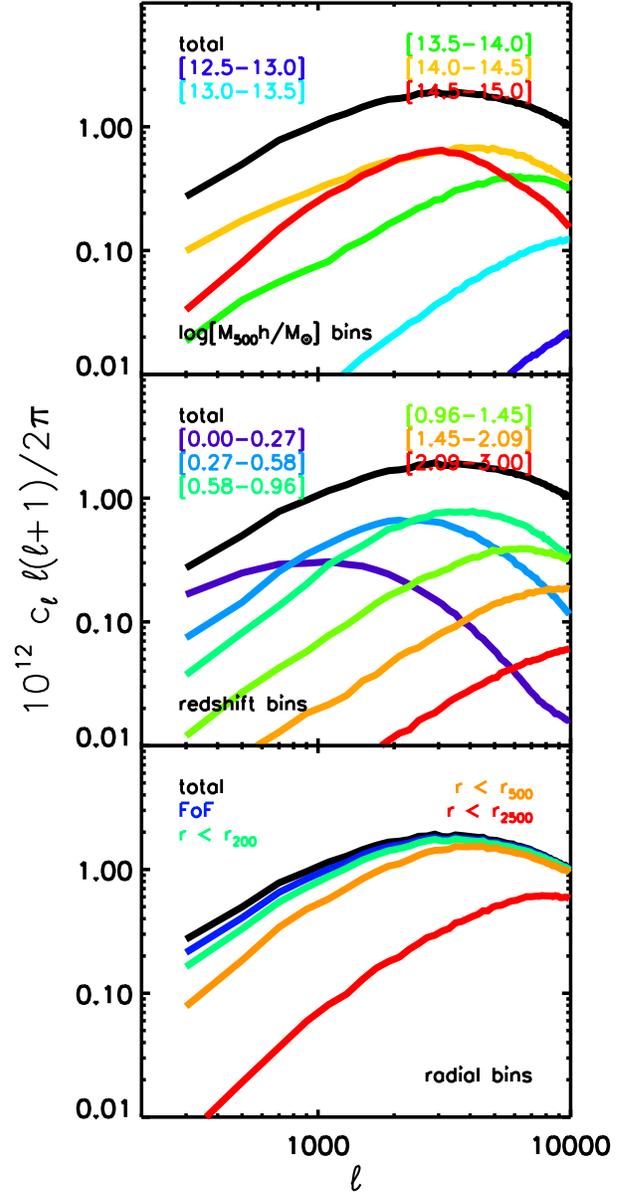}
\caption{\label{fig:cuts}
Deconstruction of the tSZ angular power spectrum.  Shown is the contribution from hot gas in haloes in bins of $\log[M_{500}h/~\textrm{M}_{\odot}]$ ({\emph top}), redshift ({\emph middle}), and radius ({\emph bottom}) for the fiducial AGN model (\agn~8.0).  At large angular scales ($\ell \la 1000$), accessible by \planck, the power spectrum is dominated by clusters ($\log M_{500}h/M_\odot \ga 14$), nearby ($z \la 0.5$) clusters with most of the power coming from large physical scales ($r \ga r_{500}$).  At intermediate angular scales ($\ell \sim 3000$), observable with SPT and ACT, the signal is still dominated by clusters but over a much wider range of redshifts (out to $z \sim 1.5$) with most of the power coming from the radial range $r_{2500} \la r \la r_{500}$. 
}
\end{figure}

\section{Deconstructing the thermal SZ effect power spectrum}

To aid our interpretation of the comparison with observations of the tSZ power spectrum below (in Section 5), we first deconstruct the simulated tSZ power spectra into its contributions from hot gas in haloes in bins of true $M_{500}$, redshift, and radius.  The results are plotted in Fig.~\ref{fig:cuts} for the fiducial AGN model (\agn~8.0) in the \planck~best-fit cosmology.  We discuss below how these trends depend on the choice of cosmology and sub-grid physics.  Note that to reduce sampling noise in the power spectra, we have re-binned to a multipole resolution of $\Delta \ell = 200$.

We consider the break down by system mass first, plotted in the top panel of Fig.~\ref{fig:cuts}.  The coloured curves correspond to power spectra from gas within $r_{200}$ in different $M_{500}$ bins.  At large angular scales ($\ell \la 1000$), accessible by the \planck~telescope, the power spectrum is dominated by relatively massive ($\log[M_{500}h/M_\odot] > 14$) systems.  A relatively larger contribution is made from galaxy groups with [$13.5-14.0$] at intermediate angular scales ($\ell \sim 3000$) observable with SPT and ACT, but clusters still dominate the signal.  It is only when one approaches scales of an arcminute ($\ell \sim 10000$) or so that the contribution of systems with masses below $10^{14} \ h^{-1}~\textrm{M}_{\odot}$ becomes comparable to that from systems with masses above this limit.

The trends in the top panel of Fig.~\ref{fig:cuts} are very similar for the \wmap~best-fit cosmology, but with a slightly increased importance of high-mass groups [$13.5-14.0$] compare to clusters [$> 14$] on the largest scales (and low-mass groups [$13.0-13.5$] compared to high-mass groups [$13.5-14.0$] on small angular scales), due to the fact that the number density of massive haloes is significantly reduced in the \wmap~cosmology compared to the \planck~cosmology.  The trends are not particularly sensitive to the nature of the implemented sub-grid physics either; massive systems with $\log[M_{500}h/M_\odot] > 14$ dominate the power spectrum at $\ell < 5000$ for all of the models we have considered.  Our trends with system mass are similar to those reported previously by \citet{Battaglia2012}, although there are differences in detail.

In the middle panel of Fig.~\ref{fig:cuts} we consider the contribution from hot gas in different redshift bins.  We have chosen the 6 redshift bins to have approximately the same comoving length ($\sim 1$ Gpc).  At large angular scales ($\ell \la 1000$), the power spectrum is produced mainly by relatively local systems with $z \la 0.5$ but with a non-negligible contribution from gas out to $z \sim 1$.  At intermediate angular scales ($\ell \sim 3000$), on the the other hand, the signal has significant contributions from $0 \la z \la 1.5$ with the range $0.25 \la z \la 1$ providing the largest contribution.  As one pushes to smaller angular scales ($\ell \sim 10000$) local sources no longer contribute significantly while gas out to $z \sim 2$ becomes important.  

\begin{figure*}
\includegraphics[width=0.49\textwidth]{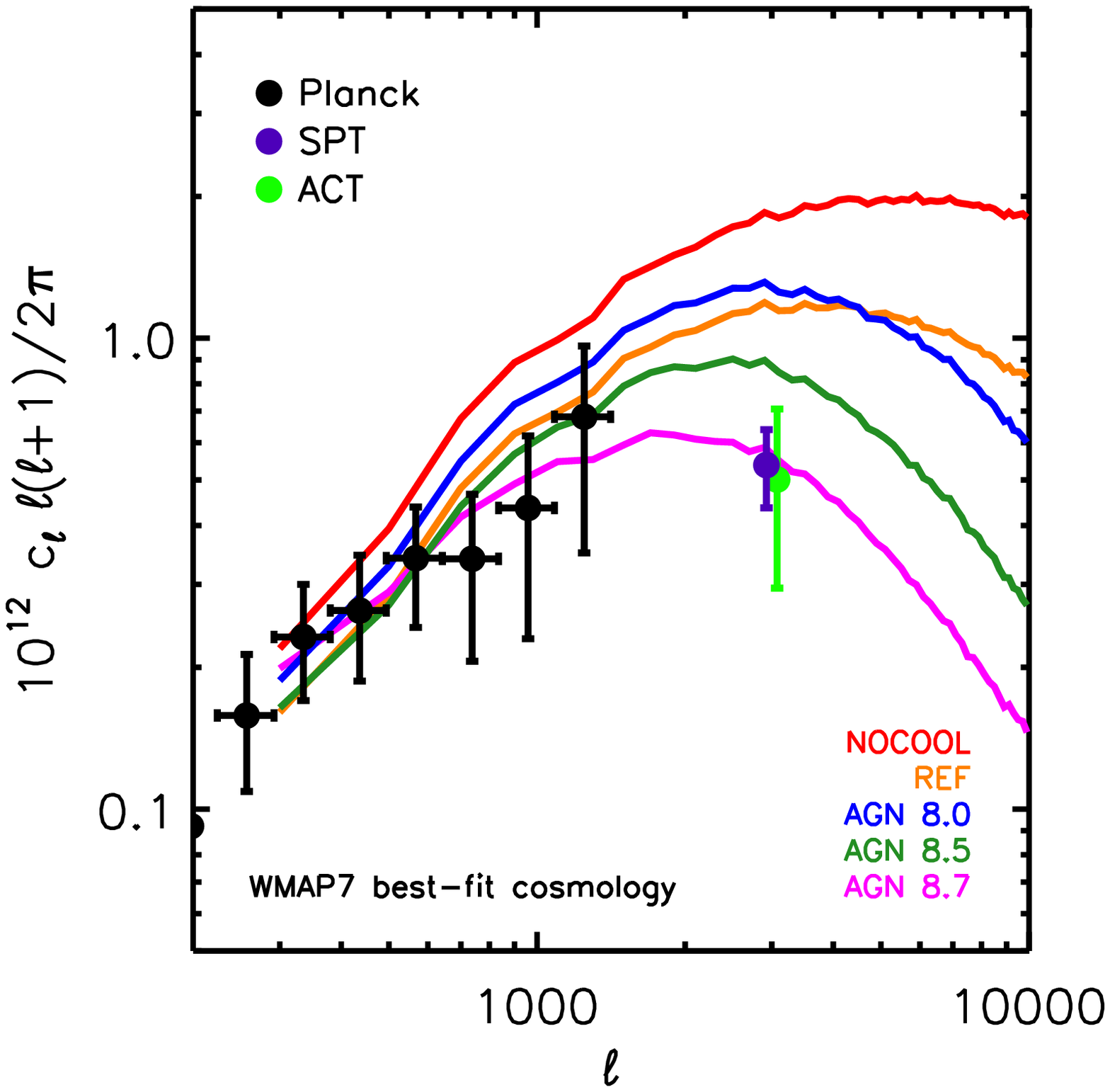}
\includegraphics[width=0.49\textwidth]{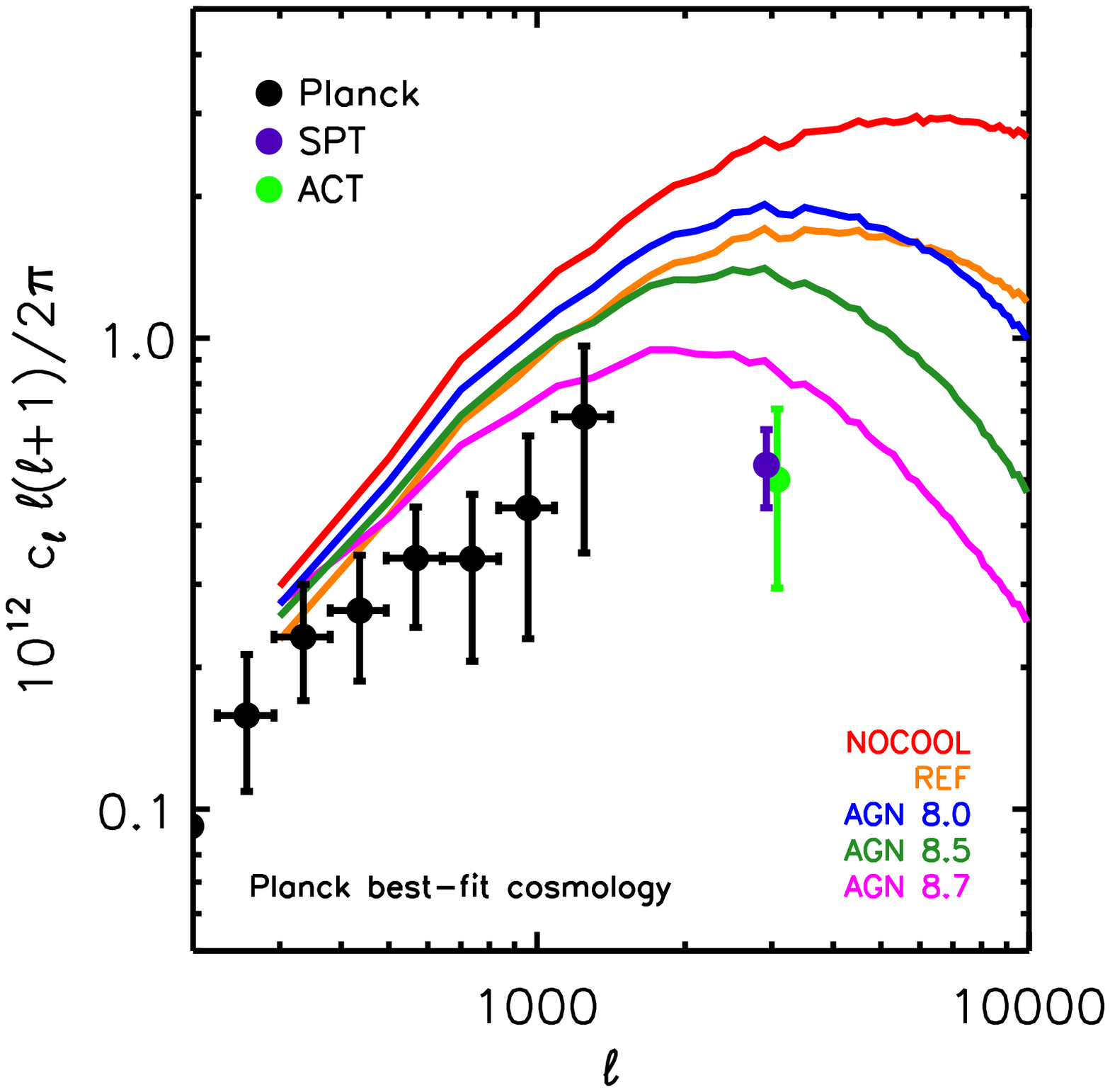}
\caption{\label{fig:data_comp}
Comparison of observed and predicted tSZ angular power spectra in the \wmap~best-fit cosmology ({\emph left}) and the \planck~best-fit cosmology ({\emph right}).  The thick coloured curves represent the mean power spectra for each of the simulations.  The filled circles with 1 sigma error bars represent measurements from the \planck~telescope (\citealt{Planck2013c}, black), the South Pole telescope (\citealt{Reichardt2012}, purple), and the Atacama Cosmology Telescope (\citealt{Sievers2013}, bright green), respectively.  There is a significant amplitude offset between the models and the observations on all angular scales in the \planck~cosmology.  In the \wmap~cosmology, by contrast, there is reasonable agreement at large angular scales.  However, the fiducial AGN model (\agn~8.0), which reproduces the properties of local groups and clusters best, has a factor of $\sim 2$ more power than observed by SPT and ACT at $\ell \approx 3000$.
}
\end{figure*}

The trends in the middle panel of Fig.~\ref{fig:cuts} are virtually identical in the \wmap~best-fit cosmology, depend only very mildly on the nature of the implemented sub-grid physics, and are similar to those reported previously by \citet{Battaglia2012}.

In the bottom panel of Fig.~\ref{fig:cuts} we compute the contribution to the power spectrum from gas in different radial ranges.  Note that $r_{200}$ is typically taken to be the virial radius and that $r_{500} \approx 0.65 r_{200}$ and $r_{2500} \approx 0.45 r_{500}$ for a NFW profile with a typical cluster concentration of $5$.  The `FoF' (blue) curve corresponds to the power spectrum from gas linked to the friends-of-friends group in which the simulated galaxy cluster lives.  This includes gas within $r_{200}$ as well as some beyond this radius.  Note that the FoF region is not constrained to be spherical, but typically $M_{\rm FoF} \sim 2 M_{200}$ (with significant scatter) for a standard linking length of 0.2 times the mean interparticle separation.

At large angular scales ($\ell \la 1000$) most of the tSZ signal comes from large physical radii, with more than half of the power coming from beyond $r_{500}$ (i.e., beyond the reach of most X-ray observations).  At intermediate angular scales ($\ell \sim 3000$) gas within the radial range $r_{2500} \la r \la r_{500}$ is the largest contributor to the power spectrum.  At angular scales of an arcminute and below ($\ell \sim 10000$), the `inner' regions ($r \la r_{2500}$) of groups and clusters begin to dominate.

The trends in the bottom panel of Fig.~\ref{fig:cuts} are virtually identical in the \wmap~best-fit cosmology for a given sub-grid model.  The fiducial AGN model has a similar behaviour to that of the \nocool~and \refsim~models.  However, increasing the AGN heating temperature boosts the contribution from gas at large radii on large angular scales ($\ell \la 1000$), due to the efficient ejection of gas from within $r_{500}$.

Comparing the trends in Fig.~\ref{fig:cuts} with the pressure profiles in Fig.~\ref{fig:pressure_profs}, our expectation is that at the large angular scales observable by the \planck~telescope ($\ell \la 1000$), the power spectrum should be relatively insensitive to sub-grid physics.  That is because these scales probe very large physical radii around relatively massive clusters.  By contrast, we should expect to find relatively large differences between the models at intermediate angular scales of $\ell \sim 3000$ (observable with SPT and ACT), since these probe intermediate radii ($r_{2500} \la r \la r_{500}$) and lower halo masses.  

\section{Comparison with observations}

In Fig.~\ref{fig:data_comp} we plot the predicted tSZ angular power spectra for the five models (thick colour curves) in both the \wmap~(left panel) and \planck~primary CMB (right panel) best-fit cosmologies, along with the latest power spectrum measurements from the \planck~telescope \citep{Planck2013c}, SPT \citep{Reichardt2012}, and ACT \citep{Sievers2013} as the data points with error bars.  Note that the observational error bars represent 1 sigma constraints on the power spectrum.  In the case of the \planck~measurements we sum the statistical and foreground uncertainties (e.g., as in Fig.~15 of \citealt{Planck2013c}).

{\it For the \planck~best-fit cosmology, a significant amplitude offset is present between all the models and the observations on all measured angular scales.}  Notably, the offset exists even at the largest angular scales, where the effects of baryon physics are minor, as can be deduced from the convergence of the models there.  By contrast, relatively good agreement is achieved at large angular scales in the \wmap~best-fit cosmology, with all but the \nocool~model being roughly consistent with the \planck~power spectrum measurements.  The lower power in the \wmap~cosmology is due primarily to the lower values of $\sigma_8$ ($0.81$ vs.\ $0.834$) and $\Omega_m$ ($0.272$ vs.\ $0.318$).

Encouragingly, these results are qualitatively consistent with the findings of \citet{Planck2013c}, who used a simple halo model analytical approach combined with the \citet{Tinker2008} mass function and the \citet{Arnaud2010} universal pressure profile to calculate a template tSZ power spectrum (see also \citealt{Efstathiou2012}).  By adjusting the amplitude of the template tSZ power spectrum, \citet{Planck2013c} derive the constraint $\sigma_8(\Omega_m/0.28)^{0.395} = 0.784 \pm 0.016$ (68\% C.~L.), which is significantly lower than inferred from the \planck~primary CMB, but is only 1 sigma lower than the \wmap~best-fit cosmology.

We point out that while there is qualitative agreement between our findings and those of the \planck~team, some quantitative differences are present.  Specifically, when we scale their best-fit halo model to the cosmology adopted in our simulations, the amplitudes of the halo model and hydrodynamical simulation power spectra differ by up to 50\% at large angular scales, in the sense that the halo model predicts more power than the hydrodynamical simulations.  As a consequence, the derived joint constraint on $\sigma_8$ and $\Omega_m$ using the halo model is roughly 1 sigma lower than what we would infer by scaling our simulations to match the observational data.  Why the \planck~halo model predicts more power than the hydrodynamical simulations at large scales (for a given cosmology) is unclear but is worth further investigation.  We note that the simple halo model approach neglects the effects of asphericity and substructure, which \citet{Battaglia2012} have demonstrated to be relevant for the tSZ power spectrum.  Furthermore, the analytic methodology neglects the relatively large intrinsic scatter in the tSZ flux of observed clusters (see, e.g., fig.~8 of L14) and assumes self-similar evolution, although the addition of scatter and alternative assumptions about evolution should be straightforward to implement.  Finally, recent simulation studies that include AGN feedback find that gas ejection can alter the halo mass function by up to $\sim$15-20\% at the massive end (e.g., \citealt{Cusworth2013,Cui2014,Velliscig2014}).  By contrast, our hydrodynamical simulations implicitly include all of these effects, which may go some ways towards explaining differences with the halo model\footnote{This is not to suggest that the halo model does not have its uses, quite the contrary; its strength lies in its ability to rapidly explore physical and cosmological parameter space, as well as probing the largest angular scales not easily accessible with self-consistent hydrodynamical simulations (e.g., \citealt{Hill2013a}).} predictions.

In spite of the relatively good agreement between the \wmap~simulations and the observations on large scales and the ability of model \agn~8.0 to simultaneously reproduce many `resolved' properties of the local group and cluster population remarkably well, the fiducial AGN model, \agn~8.0, is clearly inconsistent with the ACT and SPT measurements on intermediate angular scales ($\ell \approx 3000$).  AGN models with higher heating temperatures perform much better in this regard, but cannot be reconciled with the properties of the local group and cluster population (see L14).  

How can we interpret these findings?  One possibility is that the redshift evolution of clusters in the fiducial AGN model is not quite correct, in the sense that real clusters could have lower densities and pressures than predicted by the model at high redshift (as shown in Fig.~\ref{fig:cuts}, the power spectrum at $\ell \sim 3000$ is sensitive to high-$z$ clusters).  However, observations appear to suggest that, if anything, the gas mass fractions increase with redshift \citep{Lin2012}.  Direct comparison of the models with observations of high-redshift clusters will help clarify this question, but observational selection effects would have to be properly addressed.  

Another possibility is that the contributions from Galactic dust emission, radio galaxies, and/or the cosmic infrared background (CIB) from stellar-heated dust within galaxies (which has both clustered and unclustered components) to the ACT and SPT total power spectra have for some reason been overestimated, resulting in an underestimate of the tSZ power spectrum amplitude\footnote{This potential caveat is also applicable to the \planck~tSZ power spectrum measurements.  The dominant foreground for \planck~is thought to be the CIB and the analysis of the \planck~data adopts a prior $A_{\rm CIB} = 1 \pm 0.5$ on the amplitude of this component.  If in reality the CIB contributes negligibly to the total power, however, then this would result in a $\sim 30\%$ boost to the inferred \planck~tSZ power spectrum measurements (U.~Seljak, priv.~comm.)}.
Note that these experiments do not directly measure the tSZ power spectrum, but instead measure a total power spectrum from which the contributions of the primary CMB, radio sources, Galactic dust, the CIB, and kSZ are removed by adjusting template models for each component.

Alternatively, the \wmap~best-fit cosmology may not be quite correct.  The agreement on large scales may suggest that it is not far from the truth, but the amplitude, and to some extent the shape, of the tSZ power spectrum is very sensitive to the adopted cosmological parameters.  This, of course, is one of the primary reasons why measurements of the tSZ power spectrum are being made.
It is therefore of interest to see what the implications are of the uncertainty in the cosmological parameters for the above comparisons.  Below we employ the primary CMB Markov Chain Monte Carlo (MCMC) runs carried out by the \wmap~and \planck~teams to explore the impact of the uncertainty on the cosmological parameters inferred by \wmap~and \planck~on the predicted tSZ angular power spectrum.

\begin{figure}
\includegraphics[width=\columnwidth]{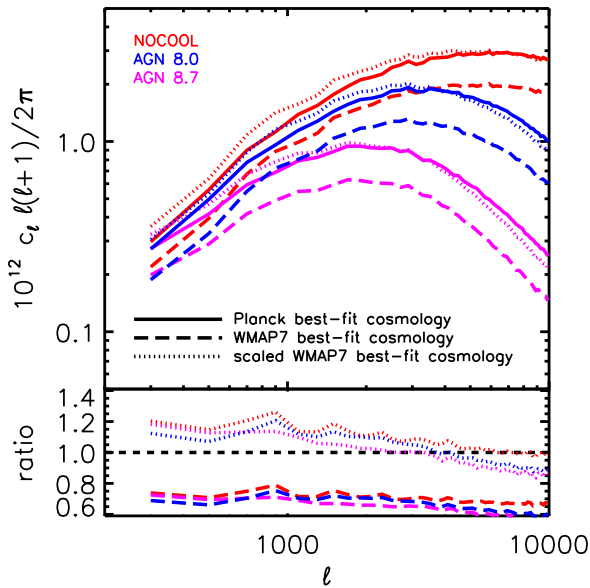}
\caption{\label{fig:scale_test}
Testing the cosmological parameter scalings proposed by \citet{Millea2012} for the tSZ power spectrum.  {\emph Top:} The solid and dashed curves represent simulations carried out with the \planck~and \wmap~best-fit cosmologies, respectively.  The dotted curves represent the \wmap~runs scaled to the \planck~cosmology using the scalings of \citet{Millea2012}.  {\emph Bottom:} The dashed curves represent the ratio of the \wmap~to \planck~cosmology runs, while the dotted curves represent the ratio of the scaled \wmap~results to the \planck~cosmology runs.  The scalings are accurate to $\approx$5-10\% at $\ell \sim 3000$ and generally accurate to $\approx$10-15\%.
}
\end{figure}

\subsection{Impact of uncertainty in cosmological parameters}

To ascertain the impact of uncertainty in the values of the cosmological parameters on the predicted tSZ power spectrum, we need a method to scale the simulated power spectra to arbitrary cosmologies (unfortunately the simulations are too expensive to run a large grid of cosmologies).  The tSZ power spectrum is most sensitive to the matter power spectrum normalisation, $\sigma_8$, but there are also relevant dependencies on the other parameters of the $\Lambda$CDM model.  To complicate things further, the relative contributions change as a function of angular scale.

\citet{Millea2012} have used the semi-analytic cluster model of \citet{Shaw2010} (which is an extension of the models originally developed by \citealt{Ostriker2005} and \citealt{Bode2009}) to construct a database of tSZ power spectra for a large grid of cosmologies.  The Shaw et al.\ model has simplified treatments of feedback due to AGN and supernovae (calibrated to reproduce the gas and stellar mass fractions of local clusters), as well as a prescription for radially-varying non-thermal pressure support\footnote{Our cosmological hydrodynamical simulations implicitly include non-thermal pressure support due to non-virialised gas.} calibrated using numerical simulations \citep{Nagai2007}.  \citet{Millea2012} fit for the dependencies of the power spectrum amplitude, as a function of multipole, of five cosmological parameters: $\Omega_{m}$, $\Omega_{b}$, $\sigma_{8}$, $n_{s}$, and $h$ ($\Omega_{\Lambda}$ is fixed by the assumption of a flat universe).  Although the derived dependencies are expected to be somewhat model dependent, they should represent an improvement over the simple $\sigma_8^n$ (where $n$ is constant $\approx 8$) scaling applied in many previous studies.

In Fig.~\ref{fig:scale_test} we test the validity of the scalings proposed by \citet{Millea2012} for our hydrodynamical simulations, by comparing how well the power spectra from our \wmap~best-fit cosmology runs agree with those from our \planck~best-fit cosmology runs when the former are scaled to the \planck~cosmology.  

As is visible from the bottom panel of Fig.~\ref{fig:scale_test}, the scalings are generally accurate to $\approx$10-15\%.  They perform slightly better than average at $\ell \sim 3000$ ($\approx$5-10\% accuracy), which is where the discrepancy between the simulations and observations is largest.  Bearing this accuracy in mind, we now proceed to use the scalings of \citet{Millea2012} to quantify the uncertainty in the predicted tSZ power spectrum due to uncertainties in cosmological parameters constrained by primary CMB measurements.

\begin{figure*}
\includegraphics[width=0.49\textwidth]{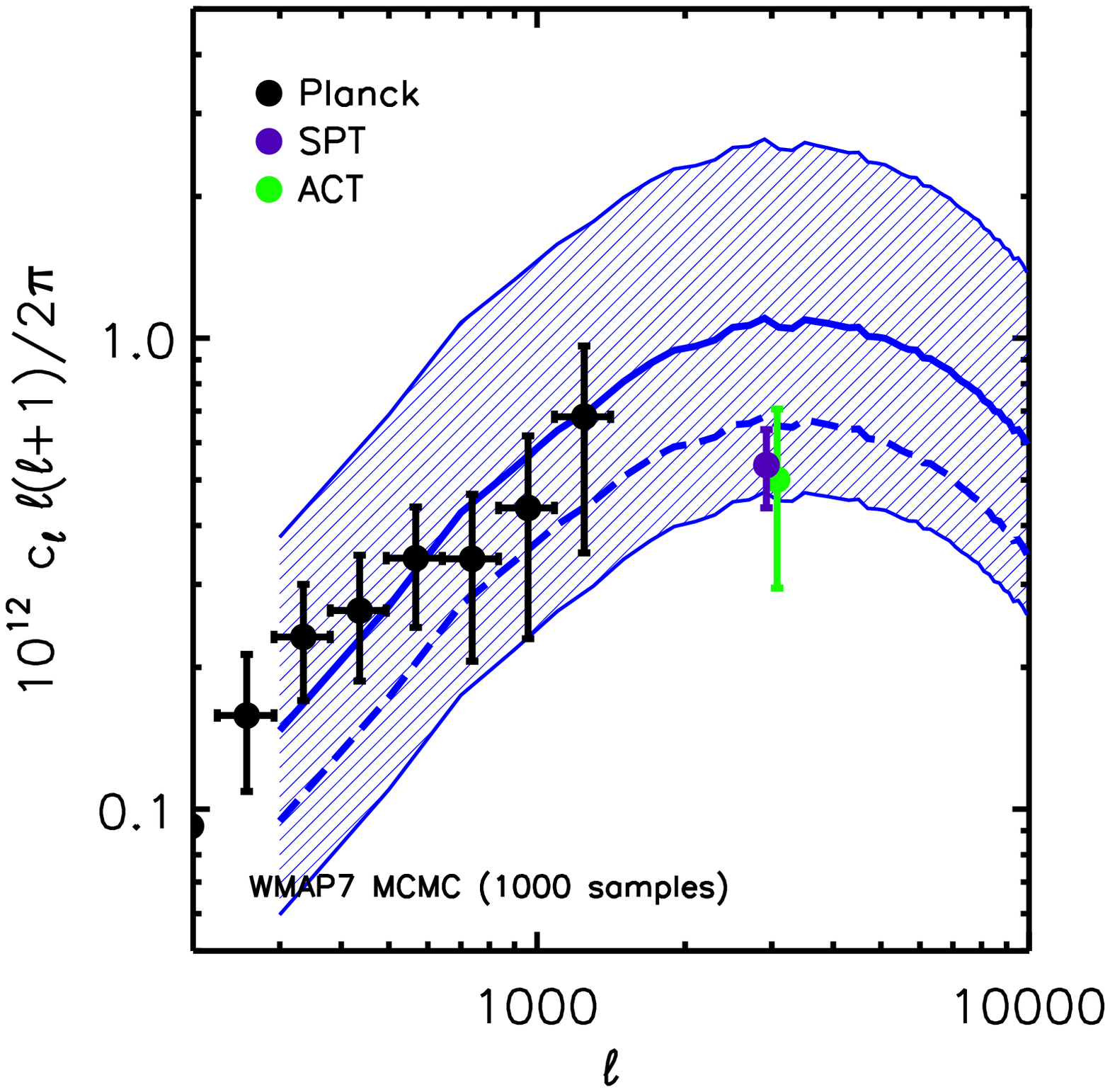}
\includegraphics[width=0.49\textwidth]{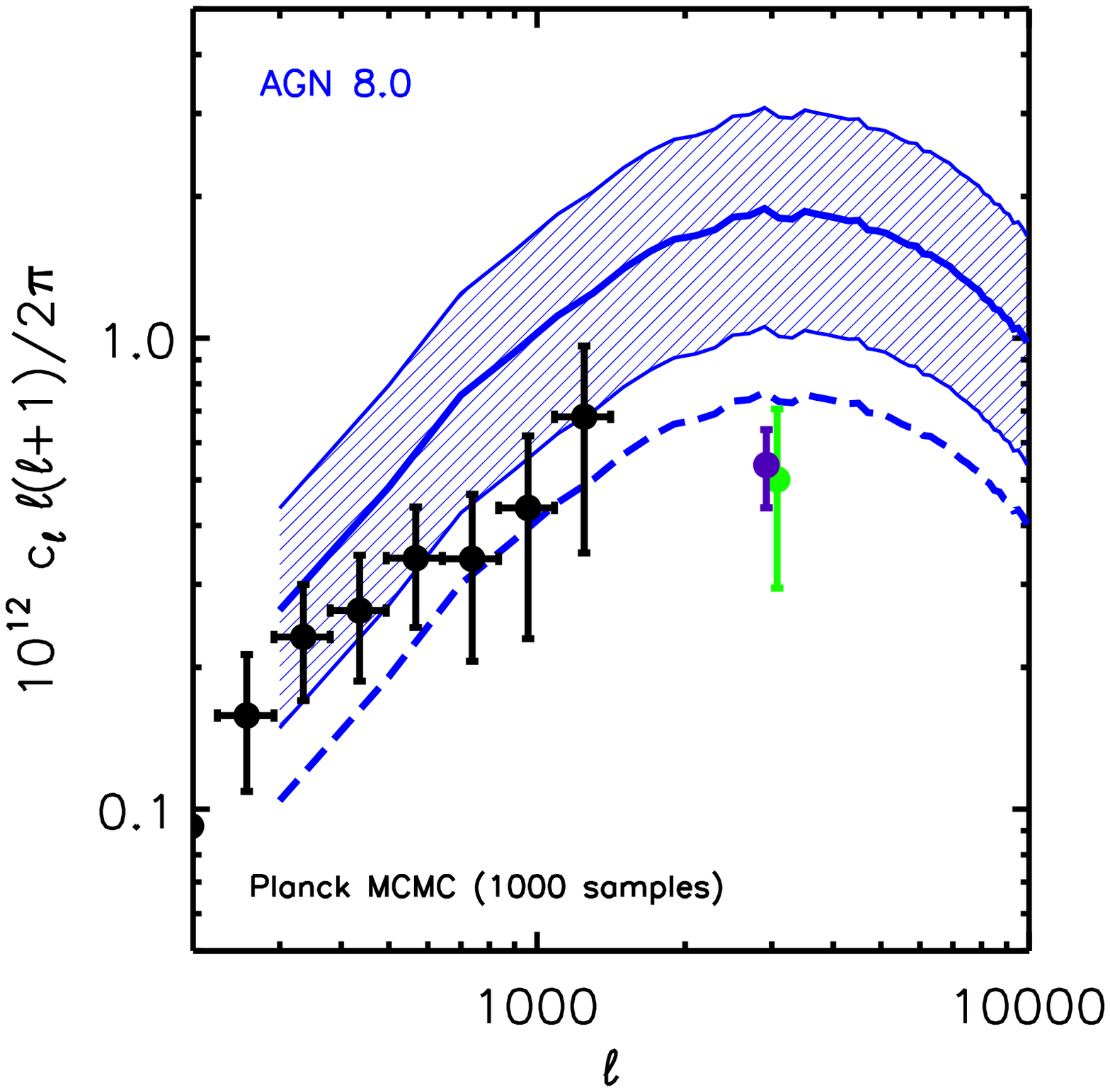}
\caption{\label{fig:mcmc_comp}
Impact of primary CMB cosmological parameter uncertainties on the predicted tSZ angular power spectrum, using the fiducial AGN model (\agn~8.0).  {\emph Left:} Using the \wmap~MCMCs.  {\emph Right:} Using the \planck~MCMCs.  The shaded region represents the uncertainty in the predicted tSZ power spectrum given the range of cosmologies allowed by the \wmap~and \planck~primary CMB constraints.  The shaded region encloses 95\% (2 sigma) of the distribution (power spectrum at a given multipole) and the thick solid curve represents the median relation.  The thick dashed curve represents the chain that gives the best match to the observed $C_\ell$ data.  The filled circles with 1 sigma error bars represent measurements from the \planck~telescope (black), the SPT (purple), and the ACT (bright green), respectively.  The predictions of the fiducial AGN model, \agn~8.0, are highly inconsistent with the \planck~and the SPT and ACT measurements using the range of $\Lambda$CDM models allowed by the \planck~primary CMB analysis.  They are, however, fully consistent with the \planck~power spectrum measurements and are consistent with the SPT and ACT data at the $\sim$2 sigma level using the range of cosmological models allowed by the \wmap~primary CMB analysis.
}
\end{figure*}

We sample the MCMC data\footnote{Publicly available on the \wmap~and \planck~websites.} produced by the \wmap~and \planck~teams, randomly selecting 1000 sets of cosmological parameter values from each.  For a given set of parameter values we use the scalings proposed by \citet{Millea2012} to adjust the tSZ power spectrum predicted by the fiducial AGN model.  We thus construct 1000 power spectra for the model for both the \wmap~and \planck~cases.  In Fig.~\ref{fig:mcmc_comp} we plot the range of power spectra that is allowed (2 sigma confidence region) by the \wmap~and \planck~primary CMB data; i.e., we have propagated the uncertainties in the primary CMB cosmological parameters to an uncertainty in the predicted tSZ angular power spectrum.

In terms of the \planck~primary CMB constraints, the predicted tSZ power spectrum is consistent with individual \planck~power spectrum measurements ($\ell \la 1000$) at the $\approx2$ sigma level (each).  How large the discrepancy is with the data set as a whole depends on the degree of covariance between neighbouring $C_\ell$'s for both the observational data and the theoretical predictions.  We note that the \planck~team have binned their data so as to minimize the covariance between neighbouring points at large scales and to maximize the signal-to-noise ratio at small scales.  Without reproducing their analysis methods exactly, it is difficult to precisely deduce the level of the discrepancy with the dataset but a (likely overly) conservative lower limit is 2 sigma.  Given the \planck~primary CMB constraints, the predicted power spectra are obviously highly inconsistent with the ACT and SPT measurements at $\ell \approx 3000$.  

The constraints placed by the \wmap~primary CMB analysis, on the other hand, are fully consistent with the \planck~power spectrum measurements and also consistent with the ACT and SPT measurements at the 2 sigma level (see shaded region).  We note, however, that no single set of parameter values (i.e., no individual chain) yields a formally good fit to the \planck, SPT, and ACT data\footnote{We have neglected the covariance between the \planck~data points for this comparison.} simultaneously in the context of the fiducial AGN model (see dashed curve in the left panel of Fig.~\ref{fig:mcmc_comp}).  The best-fit set of parameter values, obtained for the case $\sigma_8 \approx 0.766$ and $\Omega_m \approx 0.280$, has a reduced $\chi^2 \approx 2.2$.  Adopting the \agn~8.5 model does result in a formally acceptable fit (with a reduced $\chi^2 \approx 1.2$ for $\sigma_8 \approx 0.777$ and $\Omega_m \approx 0.289$) but, as already discussed, this model is in some tension with the observed properties of local groups and clusters.

In summary, if we adopt the range of $\Lambda$CDM models allowed by the \wmap~primary CMB data, we conclude that it is possible to construct a model that is consistent with the tSZ power spectrum measurements on large scales, and within 2 sigma of the data on intermediate scales, as well as with the known `resolved' properties of local groups and clusters.  By contrast, the predicted tSZ power spectra are inconsistent with the power spectrum measurements on large and intermediate scales when the range of $\Lambda$CDM models allowed is constrained by the \planck~primary CMB data.
To reconcile the \planck~primary CMB constraints with the observed power spectrum measurements requires there to be either a very different evolution in the cluster population in the models compared to reality\footnote{We have experimented with adjusting the amplitude of the deconstructed power spectra in bins of redshift for the fiducial AGN model in the \planck~best-fit cosmology.  To match the \planck~measurements at large angular scales requires a factor of $\approx3$ suppression at redshifts $z \ga 0.25$, while to match the SPT/ACT measurements requires another factor of $\approx2$ suppression for sources with $z \ga 0.5$ (so a total of $\approx6$).  Assuming the gas remains at approximately the virial temperature, this implies gas mas reductions of $\approx\sqrt3$ and $\approx\sqrt6$, respectively, for haloes with masses of $\ga 10^{14}~\textrm{M}_{\odot}$.  The former requirement appears to conflict with direct observational measurements \citep{Lin2012}.}, a departure from standard $\Lambda$CDM, or else that the tSZ power spectrum measurements are significantly biased low, e.g., due to an overestimate of the contribution of Galactic dust, radio galaxies, or the CIB to the total power spectrum.

\section{Comparison to previous studies}

Many previous theoretical studies have examined the tSZ power spectrum (e.g., \citealt{Springel2001,daSilva2001,White2002,Komatsu2002,Roncarelli2006,Holder2007,Sehgal2010,Shaw2010,Battaglia2010,Battaglia2012}).  Generally speaking, models developed prior to the first measurements of the tSZ power spectrum (by the SPT) predicted powers significantly higher than were later observed, even when the adopted cosmology was consistent with {\it WMAP} constraints.  This is likely a result of many of these early models neglecting efficient feedback from AGN, which is necessary to reconcile the models with the observed low gas densities of groups and clusters (e.g., \citealt{Puchwein2008,Bower2008,McCarthy2010}).  As we have shown above (Fig.~\ref{fig:data_comp}), such gas ejection can strongly reduce the amplitude of the predicted tSZ power spectrum on intermediate angular scales.

\begin{figure*}
\includegraphics[width=0.49\textwidth]{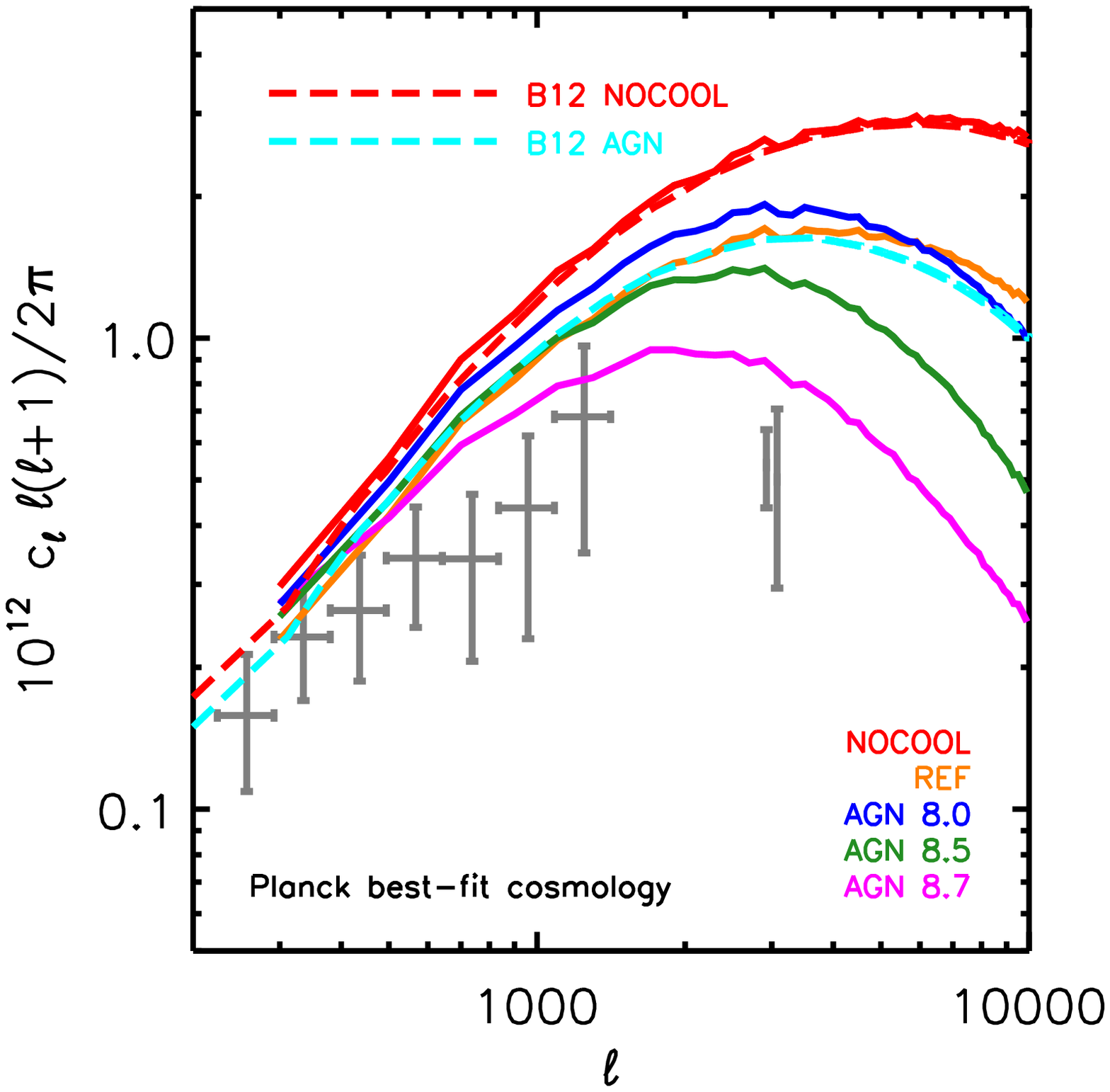}
\includegraphics[width=0.49\textwidth]{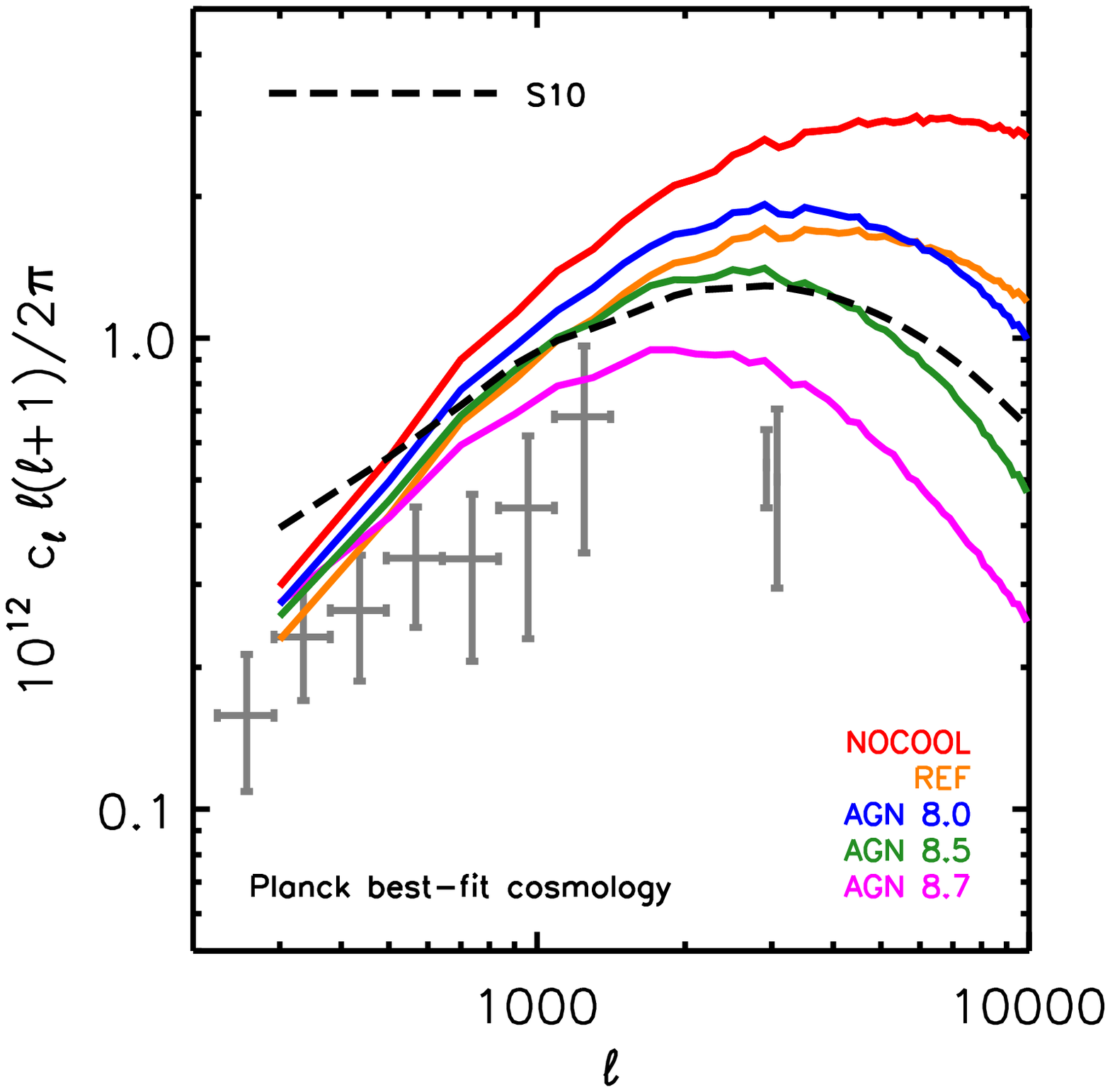}
\caption{\label{fig:sim_comp}
Comparison with the tSZ power spectra from ({\emph left}) the SPH simulations of \citet{Battaglia2012} (B12) and ({\emph right}) from the analytical model of \citet{Shaw2010} (S10).  We have scaled the power spectra of \citet{Battaglia2012} and \citet{Shaw2010} to the \planck~best-fit cosmology using the formalism of \citet{Millea2012}.  The gray data points with 1 sigma errors bars represent the measurements from the \planck~telescope, SPT, and ACT, as in previous figures.
All of the models are highly inconsistent with the ACT and SPT power spectrum measurements if the \planck~best-fit cosmology is assumed.  Note the excellent consistency of the B12 and all of our simulations at large angular scales.
}
\end{figure*}

Two of the more recent studies which have included energy input from a central engine are \citet{Shaw2010} and \citet{Battaglia2010,Battaglia2012}, with both predicting a tSZ power spectrum in approximate consistency with the SPT and ACT measurements for their adopted cosmologies.  It is therefore of interest to see how our results compare with these previous studies and to see, in particular, how robust our conclusions are on the discrepancy with the \planck~best-fit cosmology that we reported above.

In the left panel of Fig.~\ref{fig:sim_comp} we compare our predicted tSZ power spectra with the non-radiative and AGN feedback models of \citet{Battaglia2012}.  Encouragingly, there is excellent consistency between the non-radiative simulations of these authors and our own \nocool~model.  Interestingly, their AGN model predicts a tSZ power spectrum that is similar to our fiducial AGN model.  This is understandable at large angular scales (where all the simulations converge), but the agreement at intermediate and small angular scales is a bit surprising at first sight, given the sensitivity of these scales to non-gravitational physics.  It is surprising because there are large differences in the sub-grid implementations of radiative cooling (they assume primordial cooling only, whereas our simulations include metal-line cooling computed on an element-by-element basis) and AGN feedback (their feedback scales with the integrated star formation rate of their haloes, whereas ours scales with the local Bondi accretion rate), both of which can change the {\it qualitative} properties of groups and clusters \citep{McCarthy2011}.  The similarity may be tied to the fact that the AGN model of \citet{Battaglia2012} was tuned to match the gas and stellar mass fractions of a higher-resolution zoomed simulation run with the more detailed AGN model of \citet{Sijacki2008}, which, similar to our fiducial model, reproduces the baryon fractions of local groups and clusters reasonably well.

The consistency between all the simulations at large angular scales, independent of sub-grid physics, bodes well for the use of this region of the power spectrum for cosmological purposes.  It also indicates that the box sizes of current cosmological hydrodynamical simulations are sufficiently large to capture the power on these angular scales (note that the B12 simulations have box sizes of $165 \ h^{-1}$ Mpc, compared to the $400 \ h^{-1}$ Mpc boxes used here).

In the right panel of Fig.~\ref{fig:sim_comp} we compare to the semi-analytic model of \citet{Shaw2010}.  The \citet{Shaw2010} model combines a simple model for the re-distribution of the hot gas due to star formation, feedback from AGN, and radially-varying non-thermal pressure support.  The star formation efficiency and feedback parameters are tuned to match some of the properties of the local group and cluster population, while the non-thermal pressure support is constrained using cosmological simulations.  This cluster physics prescription is combined with the mass function of \citet{Tinker2008} to make predictions for the tSZ power spectrum.  Their model predicts a power spectrum that is similar to our \agn~8.5 model at intermediate and small angular scales.  There is puzzling offset from the simulation-based power spectra at large angular scales, whose origin is unclear.  These scales probe large physical radii (beyond the virial radius), suggesting the difference may be due to departures from spherical symmetry and/or an increasing importance of substructure, which are absent in the halo model approach.  Alternatively, it may signal an issue in their parameterisation of the contribution of non-thermal pressure support, which becomes increasingly important at large radii.

Overall, from Fig.~\ref{fig:sim_comp} we conclude that {\it none} of the current tSZ power spectrum predictions are consistent with the \planck~and (particularly) the SPT and ACT measurements if the \planck~best-fit cosmology is adopted.

\section{Summary and Discussion}

We have employed the cosmo-OWLS suite of large-volume cosmological hydrodynamical simulations (described in detail in \citealt{LeBrun2014}) to explore the astrophysical and cosmological dependencies of the thermal Sunyaev-Zel'dovich effect (tSZ) power spectrum.  cosmo-OWLS is an extension of the OverWhelmingly Large Simulations project \citep{Schaye2010} and has been designed specifically to aid the interpretation and analysis of cluster cosmology and large scale structure surveys.  

From the analysis presented here, we arrive at several important conclusions:

\begin{itemize}
\item{For a given cosmology, the tSZ signal on intermediate and small scales ($\ell \ga 1000$) is highly sensitive to important sub-grid physics (Fig.~\ref{fig:data_comp}), owing to the fact this range of scales probes intermediate radii in clusters (Fig.~\ref{fig:cuts}) which are susceptible to non-gravitational processes such as gas ejection due to AGN feedback (e.g., \citealt{McCarthy2011}). However, at larger scales ($\ell \ll 1000$), which probe gas at large physical radii around nearby relatively massive clusters, the effects of `sub-grid' physics are minor.}
\item{For a given physical model, the tSZ signal on all accessible scales is very sensitive to cosmological parameters that affect the abundance of the clusters, particularly $\sigma_8$ and $\Omega_m$.  Given the insensitivity of the signal to non-gravitational physics at large angular scales, this likely represents the best regime for deriving cosmological constraints.}
\item{{\it We find a significant amplitude offset between all the simulations and the observations of the tSZ power spectrum on all measured angular scales, if the \planck~best-fit cosmology is assumed by the simulations, with the simulations predicting more power than is observed} (Figs.~\ref{fig:data_comp} and \ref{fig:mcmc_comp}, right panel).  This includes the large angular scales probed by the \planck~satellite, which are insensitive to assumptions about sub-grid physics.  Note also that one of the models, the fiducial AGN model (\agn~8.0), reproduces the global X-ray, tSZ, optical, and BH scaling relations (see \citealt{LeBrun2014}), as well as the observed pressure distribution of the hot gas (Fig.~\ref{fig:pressure_profs}) of the local group and cluster population.}
\item{By contrast, if the \wmap~cosmology is adopted by the simulations, there is full consistency with the \planck~power spectrum measurements on large scales and agreement at the 2 sigma level for SPT and ACT measurements of the power spectrum at intermediate scales for the fiducial AGN model (Figs.~\ref{fig:data_comp} and \ref{fig:mcmc_comp}, left panel).  We note, however, that no single set of cosmological parameter values (in a standard 6-parameter $\Lambda$CDM model) yields a formally acceptable fit to the \planck, SPT, and ACT data simultaneously using our fiducial AGN model.}
\item{In the \wmap~cosmology it is possible to match the SPT and ACT measurements by making the AGN feedback more violent and bursty than in the fiducial AGN model (Fig.~\ref{fig:data_comp}, left panel), but this comes at the expense of spoiling the excellent agreement with the `resolved' properties of local clusters \citep{LeBrun2014}.}
\item{To reconcile the \planck~primary CMB constraints with the observed power spectrum (particularly the ACT and SPT measurements), there would have to be either a very different evolution in the cluster population in the models compared to reality (such that real clusters must be significantly under-dense/under-pressurized compared to the models at high-$z$, but observations suggest otherwise; \citealt{Lin2012}), a departure from standard $\Lambda$CDM, or else that the tSZ power spectrum data are significantly biased low, e.g., due to an overestimate of the contribution of Galactic dust, radio galaxies, or the cosmic infrared background to the total power spectrum.}
\item{By comparing our results with previous theoretical studies (namely \citealt{Shaw2010,Battaglia2012}), we show that the above conclusions are generic to current models.}
\end{itemize}

The simplest interpretation of our findings is that the lower-than-expected amplitude of the tSZ power spectrum indicates that there are significantly fewer massive dark matter haloes than expected for the \planck~primary CMB cosmology.  Indeed, \citet{Planck2013c} placed constraints on $\sigma_8$ and $\Omega_m$ using a simple halo model based approach to the tSZ power spectrum and concluded there was tension with the values derived from the primary CMB.  Interestingly, they noted that the derived constraints were fully consistent with those obtained from the tSZ cluster number counts in \citet{Planck2013b}.  In spite of this consistency, \citet{Planck2013c} suggest that the discrepancy with the primary CMB constraints is likely tied to systematics in the cluster modelling which affects both the number counts and power spectrum analyses, but in different ways.  For example, if the hydrostatic mass bias is significantly larger than currently thought, this would have the effect of lowering the number of haloes above a given tSZ flux.  At the same time, this mass bias would introduce an error in the halo modelling approach of the power spectrum, since it adopts empirical constraints between the tSZ flux signal and halo mass (namely the universal pressure profile of \citealt{Arnaud2010}).

However, we have shown here that a significant discrepancy exists in the amplitude of the predicted and observed tSZ power spectrum that does not rely on the tSZ flux-mass relation being known, and also addresses other criticisms of the halo model approach (e.g., asphericity, substructure, intrinsic scatter, a halo mass function that includes modifications due to baryons).  We simply compare the power spectra of simulated and observed tSZ skies, where the simulated tSZ skies are produced from fully self-consistent cosmological hydrodynamical simulations including one that reproduces optical and X-ray observations of local groups and clusters.  We point out that our results are consistent with other \planck~tSZ-derived constraints on $\sigma_8$ and $\Omega_m$, including those derived from the cross-correlation of X-ray clusters \citep{Haijan2013} and CMB lensing \citep{Hill2013b} with the \planck~tSZ signal, as well as constraints from galaxy-galaxy lensing and galaxy clustering (e.g., \citealt{Cacciato2013}).

\begin{figure}
\includegraphics[width=\columnwidth]{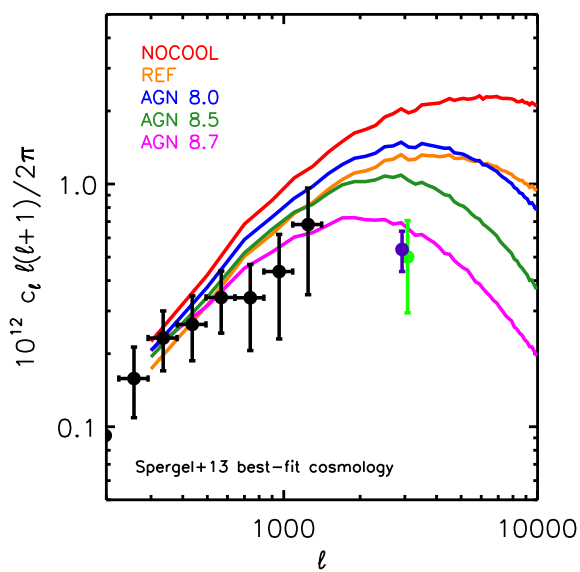}
\caption{\label{fig:spergel_test}
The tSZ power spectra in a \citet{Spergel2013} best-fit cosmology.  The amplitude offset with respect to the data is now largely removed and predicted spectra are very similar to that obtained in \wmap~best-fit cosmology (see the left panel of Fig.~\ref{fig:data_comp}).
}
\end{figure}

At face value, therefore, our results pose a significant challenge to the cosmological parameter values preferred (and/or the model adopted) by the \planck~primary CMB analyses. To be definitive, however, confirmation of these findings using other simulations is needed.  Furthermore, a more rigorous comparison between the simulated tSZ skies and observations should be undertaken, by bringing the simulated tSZ skies fully to the observational plane (instrumental response + noise + contamination) and then analysing them using the same pipeline as used on the real data.  In addition, since the power spectrum is sensitive to high-$z$ clusters, it will be important to confront the models with resolved observations of such systems (but care must be taken to address important observational selection effects).

While finalising this paper a re-analysis of the \planck~primary CMB data by \citet{Spergel2013} was posted to the arXiv.  These authors claim to have identified a systematic issue with the 217 GHz $\times$ 217 GHz detector set spectrum used in the \planck~analysis.  When corrected for, \citet{Spergel2013} find that some of the tension between the \planck~best-fit parameters and previous cosmological constraints is removed.  We have used the best-fit cosmological parameters derived by \citet{Spergel2013} to see what impact this has on the predicted tSZ power spectrum.  We scale the simulated power spectra to the Spergel et al.\ best-fit cosmology.  The results are shown in Fig.~\ref{fig:spergel_test} and show that the amplitude offset is significantly reduced for this revised cosmology.
 
\section*{Acknowledgements}

The authors would like to thank the referee, Stefano Borgani, for useful suggestions which improved the quality of the paper.  They also thank the members of the OWLS team for their contributions to the development of the simulation code used here, as well as Nick Battaglia, George Efstathiou, Colin Hill, Scott Kay, Eiichiro Komatsu, Uros Seljak, David Spergel, and Simon White for helpful discussions.  Further thanks to Nick Battaglia for providing his simulated power spectra and Ming Sun for providing his pressure profile measurements.  
IGM is supported by an STFC Advanced Fellowship at Liverpool John Moores University. AMCLB acknowledges support from an internally funded PhD studentship at the Astrophysics Research Institute of Liverpool John Moores University.  JS is sponsored by the European Research Council under
the European Union's Seventh Framework Programme (FP7/2007-2013)/ERC Grant agreement 278594-GasAroundGalaxies.  This work used the DiRAC Data Centric system at Durham University, operated by the Institute for Computational Cosmology on behalf of the STFC DiRAC HPC Facility (www.dirac.ac.uk). This equipment was funded by BIS National E-infrastructure capital grant ST/K00042X/1, STFC capital grant ST/H008519/1, and STFC DiRAC Operations grant ST/K003267/1 and Durham University. DiRAC is part of the National E-Infrastructure.

\appendix
\section{Resolution study}

\begin{figure}
\includegraphics[width=\columnwidth]{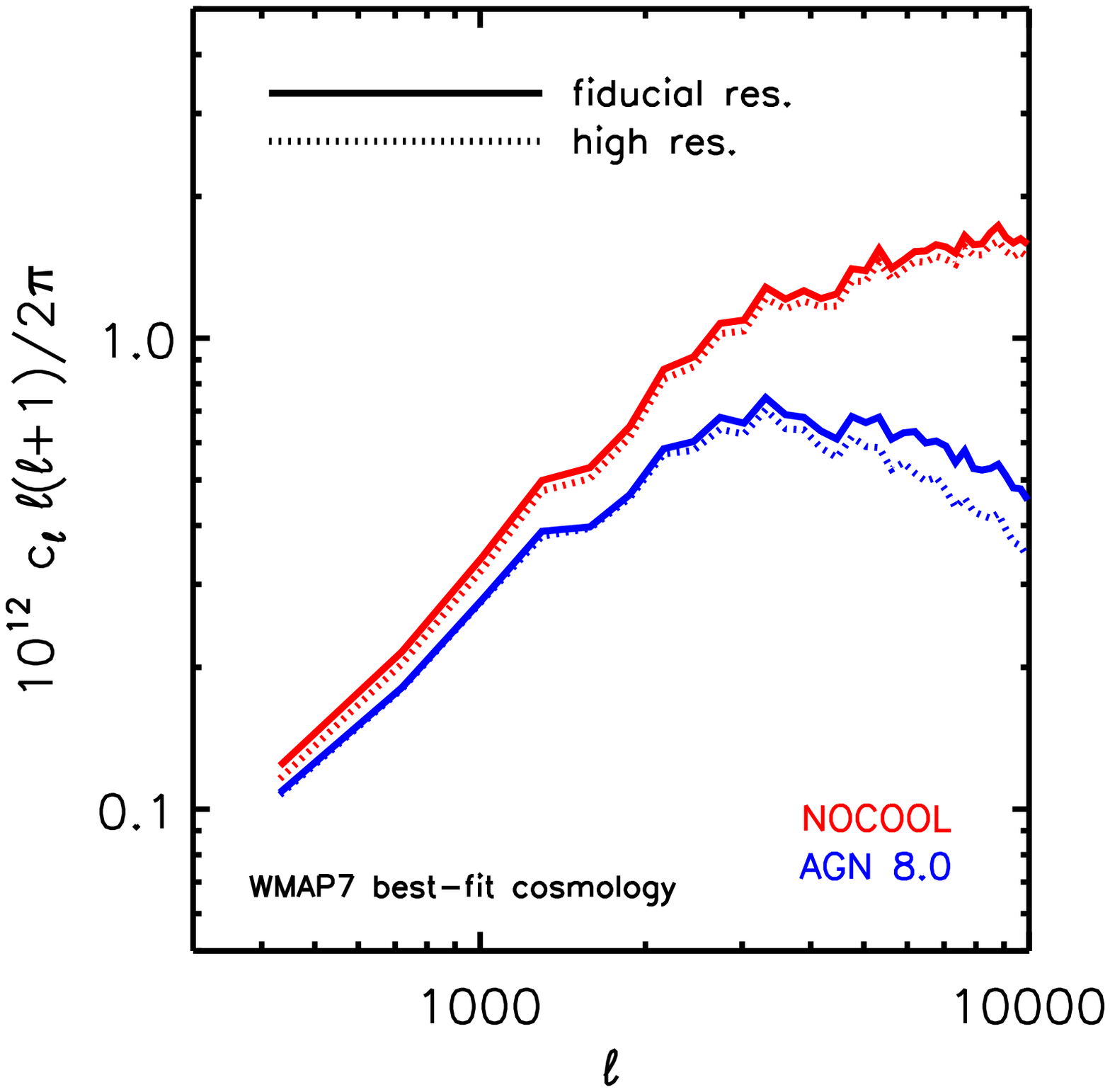}
\caption{\label{fig:res_test}
Predicted tSZ power spectra for the fiducial and high-resolution simulations.  These are based on 1.25 deg.\ $\times$ 1.25 deg.\ surveys constructed from $100 \ h^{-1}$ Mpc simulations with $256^3$ (fiducial) and $512^3$ (high-res.) baryon and dark matter particles.  The tSZ power spectra are reasonably well converged at all angular scales in the \nocool~run and at $\ell \la 4000$ for the fiducial AGN run.  
}
\end{figure}

In Fig.~\ref{fig:res_test} we present a numerical resolution convergence study for the predicted tSZ power spectra.  For this test we use $100 \ h^{-1}$ Mpc box simulations with $256^3$ (fiducial) and $512^3$ (high-res.) baryon and dark matter particles.  The latter has a factor of $8$ ($2$) better mass (spatial) resolution than the former.  (Note that a $400 \ h^{-1}$ Mpc box with $1024^3$ particles has the same resolution as a $100 \ h^{-1}$ Mpc box with $256^3$ particles.)  The smaller box size imposes a smaller field of view, we thus construct 1.25 deg.\ $\times$ 1.25 deg.\ lightcones back to $z=3$.  

The tSZ power spectra are well-converged at all angular scales in the \nocool~run and at $\ell \la 4000$ for the fiducial AGN run.  At $\ell = 10000$ the high-res.\ simulation has approx. 25\% less power compared to the fiducial run.

\section{Cosmic variance}

\begin{figure}
\includegraphics[width=\columnwidth]{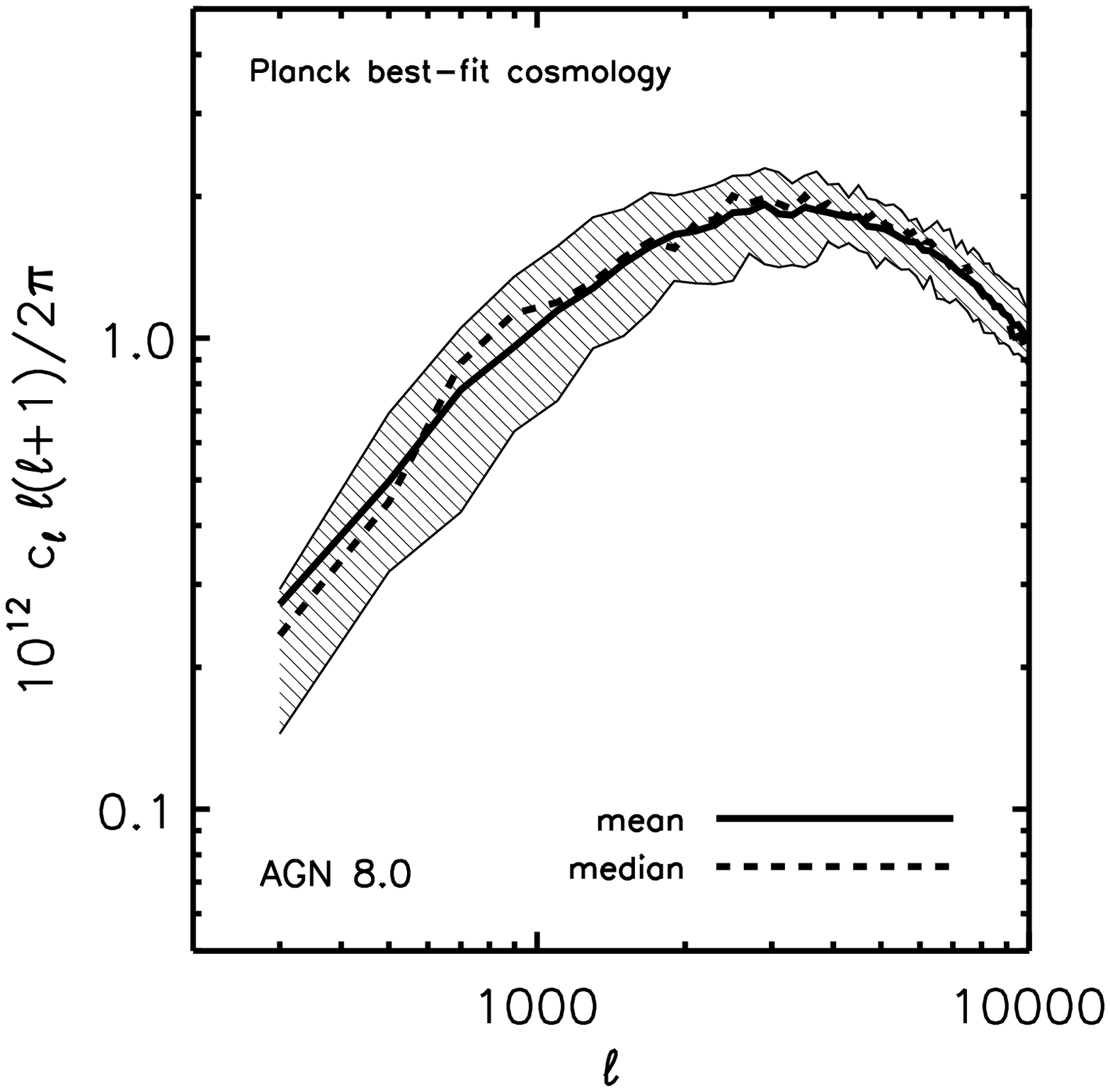}
\caption{\label{fig:variance_test}
Map-to-map scatter in the predicted tSZ power spectrum for the fiducial AGN model for the 25 degree$^2$ maps.  The solid and dashed curves represent the mean and median power spectra from the 10 lightcone realizations.  The shaded region encloses the 10th and 90th percentiles.  At large angular scales ($\ell \la 1000$) the scatter in the power spectrum can reach $\sim 50\%$, while at small scales ($\ell \ga 3000$) it is typically $\sim 20\%$.
}
\end{figure}

In Fig.~\ref{fig:variance_test} we show the map-to-map scatter in the predicted tSZ power spectra for the fiducial AGN model in the \planck~best-fit cosmology.  The 
shaded region encloses the 10th and 90th percentiles.  The scatter can reach up to $\sim 50\%$ at large angular scales ($\ell \la 1000$), but is typically only  $\sim 20\%$ at intermediate/small scales ($\ell \ga 3000$).  

We stress that such cosmic variance is likely negligible for current large observational surveys which have areas of hundreds and thousands of square degrees.  It is also negligible for simple halo model calculations which can probe arbitrarily large volumes.  For self-consistent cosmological hydrodynamical simulations, such as the ones presented in this paper, it is crucial that the volume is sufficiently large (and/or the number of independent volumes analysed is sufficiently large) to be able to robustly estimate the true mean tSZ power spectrum on the range of scales of interest.  The fact that there is excellent agreement between our mean power spectra at large angular scales, which were extracted from the same simulation and are thus not independent, and that of the simulations of \citet{Battaglia2012} (see the left panel of Fig.~\ref{fig:sim_comp}), who use 10 independent but smaller volumes, indicates that current hydrodynamical simulations are sufficiently large to measure the mean power spectrum accurately.


\begin{thebibliography}{99}

\bibitem[\protect\citeauthoryear{Arnaud et al.}{2010}]{Arnaud2010} Arnaud M., Pratt G.~W., Piffaretti R., B{\"o}hringer H., Croston J.~H., Pointecouteau E., 2010, A\&A, 517, A92 

\bibitem[\protect\citeauthoryear{Battaglia et al.}{2010}]{Battaglia2010} Battaglia N., Bond J.~R., Pfrommer C., Sievers J.~L., Sijacki D., 2010, ApJ, 725, 91 

\bibitem[\protect\citeauthoryear{Battaglia et al.}{2012}]{Battaglia2012} Battaglia N., Bond J.~R., Pfrommer C., Sievers J.~L., 2012, ApJ, 758, 75 

\bibitem[\protect\citeauthoryear{Birkinshaw}{1999}]{Birkinshaw1999} Birkinshaw M., 1999, PhR, 310, 97 

\bibitem[\protect\citeauthoryear{Bode, Ostriker \& Vikhlinin}{2009}]{Bode2009} Bode P., Ostriker J.~P., Vikhlinin A., 2009, ApJ, 700, 989 

\bibitem[\protect\citeauthoryear{B{\"o}hringer et al.}{2007}]{Bohringer2007} B{\"o}hringer H., et al., 2007, A\&A, 469, 363 

\bibitem[\protect\citeauthoryear{Booth \& Schaye}{2009}]{Booth2009} Booth C.~M., Schaye J., 2009, MNRAS, 398, 53 

\bibitem[\protect\citeauthoryear{Bower, McCarthy \& Benson}{2008}]{Bower2008} Bower R.~G., McCarthy I.~G., Benson A.~J., 2008, MNRAS, 390, 1399 

\bibitem[\protect\citeauthoryear{Cacciato et al.}{2013}]{Cacciato2013} Cacciato M., van den Bosch F.~C., More S., Mo H., Yang X., 2013, MNRAS, 430, 767 

\bibitem[\protect\citeauthoryear{Carlstrom, Holder \& Reese}{2002}]{Carlstrom2002} Carlstrom J.~E., Holder G.~P., Reese E.~D., 2002, ARA\&A, 40, 643 

\bibitem[\protect\citeauthoryear{Cui, Borgani, 
\& Murante}{2014}]{Cui2014} Cui W., Borgani S., Murante G., 2014, MNRAS, submitted (arXiv:1402.1493)

\bibitem[\protect\citeauthoryear{Cusworth et al.}{2013}]{Cusworth2013} Cusworth S.~J., Kay S.~T., Battye R.~A., Thomas P.~A., 2013, MNRAS, submitted (arXiv:1309.4094)

\bibitem[\protect\citeauthoryear{da Silva et al.}{2000}]{daSilva2000} da Silva A.~C., Barbosa D., Liddle A.~R., Thomas P.~A., 2000, MNRAS, 317, 37 

\bibitem[\protect\citeauthoryear{da Silva et al.}{2001}]{daSilva2001} da Silva A.~C., Kay S.~T., Liddle A.~R., Thomas P.~A., Pearce F.~R., Barbosa D., 2001, ApJ, 561, L15 

\bibitem[\protect\citeauthoryear{Dalla Vecchia \& Schaye}{2008}]{DallaVecchia2008} Dalla Vecchia C., Schaye J., 2008, MNRAS, 387, 1431 

\bibitem[\protect\citeauthoryear{Efstathiou \& Migliaccio}{2012}]{Efstathiou2012} Efstathiou G., Migliaccio M., 2012, MNRAS, 423, 2492 

\bibitem[\protect\citeauthoryear{Hajian et al.}{2013}]{Haijan2013} Hajian A., Battaglia N., Spergel D.~N., Bond J.~R., Pfrommer C., Sievers J.~L., 2013, JCAP, 11, 64 

\bibitem[\protect\citeauthoryear{Hand et al.}{2012}]{Hand2012} Hand N., et al., 2012, PhRvL, 109, 041101 

\bibitem[\protect\citeauthoryear{Hill \& Pajer}{2013}]{Hill2013a} Hill J.~C., Pajer E., 2013, PhRvD, 88, 063526 

\bibitem[\protect\citeauthoryear{Hill \& Spergel}{2013}]{Hill2013b} Hill J.~C., Spergel D.~N., 2013, JCAP, in press (arXiv:1312.4525)

\bibitem[\protect\citeauthoryear{Holder, McCarthy \& Babul}{2007}]{Holder2007} Holder G.~P., McCarthy I.~G., Babul A., 2007, MNRAS, 382, 1697 

\bibitem[\protect\citeauthoryear{Jenkins et al.}{2001}]{Jenkins2001} Jenkins A., Frenk C.~S., White S.~D.~M., Colberg J.~M., Cole S., Evrard A.~E., Couchman H.~M.~P., Yoshida N., 2001, MNRAS, 321, 372 

\bibitem[\protect\citeauthoryear{Komatsu \& Kitayama}{1999}]{Komatsu1999} Komatsu E., Kitayama T., 1999, ApJ, 526, L1

\bibitem[\protect\citeauthoryear{Komatsu \& Seljak}{2002}]{Komatsu2002} Komatsu E., Seljak U., 2002, MNRAS, 336, 1256 

\bibitem[\protect\citeauthoryear{Komatsu et al.}{2011}]{Komatsu2011} Komatsu E., et al., 2011, ApJS, 192, 18 

\bibitem[\protect\citeauthoryear{Le Brun et al.}{2014}]{LeBrun2014} Le Brun A.~M.~C., McCarthy I.~G., Schaye J., Ponman T.~J., 2014, MNRAS, submitted (arXiv:1312.5462)

\bibitem[\protect\citeauthoryear{Lin et al.}{2012}]{Lin2012} Lin Y.-T., Stanford S.~A., Eisenhardt P.~R.~M., Vikhlinin A., Maughan B.~J., Kravtsov A., 2012, ApJ, 745, L3 

\bibitem[\protect\citeauthoryear{McCarthy et al.}{2010}]{McCarthy2010} McCarthy I.~G., et al., 2010, MNRAS, 406, 822 

\bibitem[\protect\citeauthoryear{McCarthy et al.}{2011}]{McCarthy2011} McCarthy I.~G., Schaye J., Bower R.~G., Ponman T.~J., Booth C.~M., Dalla Vecchia C., Springel V., 2011, MNRAS, 412, 1965 

\bibitem[\protect\citeauthoryear{Millea et al.}{2012}]{Millea2012} Millea M., Dor{\'e} O., Dudley J., Holder G., Knox L., Shaw L., Song Y.-S., Zahn O., 2012, ApJ, 746, 4 

\bibitem[\protect\citeauthoryear{Mo, van den Bosch, \& White}{2010}]{Mo2010} Mo H., van den Bosch F.~C., White S., 2010, Galaxy Formation and Evolution. Cambridge Univ. Press, Cambridge

\bibitem[\protect\citeauthoryear{Nagai, Kravtsov \& Vikhlinin}{2007}]{Nagai2007} Nagai D., Kravtsov A.~V., Vikhlinin A., 2007, ApJ, 668, 1 

\bibitem[\protect\citeauthoryear{Ostriker, Bode \& Babul}{2005}]{Ostriker2005} Ostriker J.~P., Bode P., Babul A., 2005, ApJ, 634, 964 

\bibitem[\protect\citeauthoryear{Planck Collaboration XVI}{2013}]{Planck2013a} Planck Collaboration XVI, 2013, A\&A, submitted (arXiv:1303.5076)

\bibitem[\protect\citeauthoryear{Planck Collaboration XX}{2013}]{Planck2013b} Planck Collaboration XX, 2013, A\&A, submitted (arXiv:1303.5080)

\bibitem[\protect\citeauthoryear{Planck Collaboration XXI}{2013}]{Planck2013c} Planck Collaboration XXI, 2013, A\&A, submitted (arXiv:1303.5081)

\bibitem[\protect\citeauthoryear{Planelles et al.}{2013}]{Planelles2013} Planelles S., Borgani S., Fabjan D., Killedar M., Murante G., Granato G.~L., Ragone-Figueroa C., Dolag K., 2013, MNRAS, in press (arXiv:1311.0818)

\bibitem[\protect\citeauthoryear{Puchwein, Sijacki \& Springel}{2008}]{Puchwein2008} Puchwein E., Sijacki D., Springel V., 2008, ApJ, 687, L53 

\bibitem[\protect\citeauthoryear{Reichardt et al.}{2012}]{Reichardt2012} Reichardt C.~L., et al., 2012, ApJ, 755, 70 

\bibitem[\protect\citeauthoryear{Roncarelli et al.}{2006}]{Roncarelli2006} Roncarelli M., Moscardini L., Tozzi P., Borgani S., Cheng L.~M., Diaferio A., Dolag K., Murante G., 2006, MNRAS, 368, 74 

\bibitem[\protect\citeauthoryear{Roncarelli et al.}{2007}]{Roncarelli2007} Roncarelli M., Moscardini L., Borgani S., Dolag K., 2007, MNRAS, 378, 1259 

\bibitem[\protect\citeauthoryear{Sayers et al.}{2013}]{Sayers2013} Sayers J., et al., 2013, ApJ, 778, 52 

\bibitem[\protect\citeauthoryear{Schaye \& Dalla Vecchia}{2008}]{Schaye2008} Schaye J., Dalla Vecchia C., 2008, MNRAS, 383, 1210 

\bibitem[\protect\citeauthoryear{Schaye et al.}{2010}]{Schaye2010} Schaye, J., Dalla Vecchia, C., Booth, C.~M., et al.\ 2010, MNRAS, 402, 1536 

\bibitem[\protect\citeauthoryear{Sehgal et al.}{2010}]{Sehgal2010} Sehgal N., Bode P., Das S., Hernandez-Monteagudo C., Huffenberger K., Lin Y.-T., Ostriker J.~P., Trac H., 2010, ApJ, 709, 920 

\bibitem[\protect\citeauthoryear{Shaw et al.}{2010}]{Shaw2010} Shaw L.~D., Nagai D., Bhattacharya S., Lau E.~T., 2010, ApJ, 725, 1452 

\bibitem[\protect\citeauthoryear{Short \& Thomas}{2009}]{Short2009} Short C.~J., Thomas P.~A., 2009, ApJ, 704, 915 

\bibitem[\protect\citeauthoryear{Sievers et al.}{2013}]{Sievers2013} Sievers J.~L., et al., 2013, JCAP, 10, 60 

\bibitem[\protect\citeauthoryear{Sijacki et al.}{2008}]{Sijacki2008} Sijacki D., Pfrommer C., Springel V., En{\ss}lin T.~A., 2008, MNRAS, 387, 1403 

\bibitem[Spergel et al.(2013)]{Spergel2013} Spergel, D., Flauger, 
R., \& Hlozek, R.\ 2013, arXiv:1312.3313 

\bibitem[\protect\citeauthoryear{Springel, White \& Hernquist}{2001}]{Springel2001} Springel V., White M., Hernquist L., 2001, ApJ, 549, 681 

\bibitem[\protect\citeauthoryear{Springel, Di Matteo \& Hernquist}{2005}]{Springel2005a} Springel V., Di Matteo T., Hernquist L., 2005, MNRAS, 361, 776 

\bibitem[\protect\citeauthoryear{Springel}{2005}]{Springel2005b} Springel V., 2005, MNRAS, 364, 1105 

\bibitem[\protect\citeauthoryear{Sun et al.}{2011}]{Sun2011} Sun M., Sehgal N., Voit G.~M., Donahue M., Jones C., Forman W., Vikhlinin A., Sarazin C., 2011, ApJ, 727, L49 

\bibitem[\protect\citeauthoryear{Sunyaev \& Zeldovich}{1972}]{Sunyaev1972} Sunyaev R.~A., Zeldovich Y.~B., 1972, CoASP, 4, 173 

\bibitem[\protect\citeauthoryear{Tinker et al.}{2008}]{Tinker2008} Tinker J., Kravtsov A.~V., Klypin A., Abazajian K., Warren M., Yepes G., Gottl{\"o}ber S., Holz D.~E., 2008, ApJ, 688, 709 

\bibitem[\protect\citeauthoryear{Trac, Bode \& Ostriker}{2011}]{Trac2011} Trac H., Bode P., Ostriker J.~P., 2011, ApJ, 727, 94 

\bibitem[\protect\citeauthoryear{van Daalen et al.}{2013}]{vanDaalen2013} van Daalen, M.~P., Schaye, J., McCarthy, I.~G., Booth, C.~M. \& Dalla Vecchia, C.\ 2013, MNRAS, submitted (arXiv:1310.7571)

\bibitem[\protect\citeauthoryear{Velliscig et al.}{2014}]{Velliscig2014} Velliscig M., van Daalen M.~P., Schaye J., McCarthy I.~G., Cacciato M., Le Brun A.~M.~C., Dalla Vecchia C., 2014, MNRAS, submitted (arXiv:1402.4461)

\bibitem[\protect\citeauthoryear{Voit}{2005}]{Voit2005} Voit G.~M., 2005, RvMP, 77, 207

\bibitem[\protect\citeauthoryear{White, Hernquist \& Springel}{2002}]{White2002} White M., Hernquist L., Springel V., 2002, ApJ, 579, 16 

\bibitem[\protect\citeauthoryear{Wiersma, Schaye \& Smith}{2009a}]{Wiersma2009a} Wiersma R.~P.~C., Schaye J., Smith B.~D., 2009, MNRAS, 393, 99 

\bibitem[\protect\citeauthoryear{Wiersma et al.}{2009b}]{Wiersma2009b} Wiersma R.~P.~C., Schaye J., Theuns T., Dalla Vecchia C., Tornatore L., 2009, MNRAS, 399, 574 


\end{thebibliography}
\end{document}